\documentclass[fleqn,usenatbib]{mnras}

\usepackage{newtxtext,newtxmath}

\usepackage[T1]{fontenc}
\usepackage[normalem]{ulem}  

\DeclareRobustCommand{\VAN}[3]{#2}
\let\VANthebibliography\thebibliography
\def\thebibliography{\DeclareRobustCommand{\VAN}[3]{##3}\VANthebibliography}

\usepackage{graphicx}	
\usepackage{amsmath}	
\usepackage{float}

\newcommand{\tessreduce}[1]{\texttt{TESSreduce}}

\title[TESSreduce]{TESSreduce: transient focused \textit{TESS} data reduction pipeline}

\author[R. Ridden-Harper et al.]{
R. Ridden-Harper,$^{1,2}$\thanks{E-mail:\href{mailto:ryan.ridden@canterbury.ac.nz}{ryan.ridden@canterbury.ac.nz}}
A. Rest,$^{2,3}$
R. Hounsell,$^{4,5}$
T. E. Müller-Bravo,$^6$
Q. Wang,$^2$
V. A. Villar$^{7,8,9}$
\\
$^{1}$ School of Physical and Chemical Sciences | Te Kura Mat\={u}, University of Canterbury,
Private Bag 4800, Christchurch 8140,
New Zealand\\
$^{2}$ Department of Physics and Astronomy, The Johns Hopkins University, Baltimore, MD 21218, USA.\\
$^{3}$ Space Telescope Science Institute, 3700 San Martin Drive, Baltimore, MD 21218, USA.\\
$^{4}$ University of Maryland, Baltimore County, 1000 Hilltop Cir, Baltimore, MD 21250, USA\\
$^{5}$ NASA Goddard Space Flight Center, Greenbelt, MD 20771, USA\\
$^{6}$ School of Physics and Astronomy, University of Southampton, Southampton, Hampshire, SO17 1BJ, UK\\
$^{7}$ Department of Astronomy \& Astrophysics, The Pennsylvania State University, University Park, PA 16802, USA\\
$^{8}$ Institute for Computational \& Data Sciences, The Pennsylvania State University, University Park, PA 16802, USA\\
$^{9}$ Institute for Gravitation and the Cosmos, The Pennsylvania State University, University Park, PA 16802, USA\\
}
\date{Accepted XXX. Received YYY; in original form ZZZ}

\pubyear{2015}

\begin{document}
\label{firstpage}
\pagerange{\pageref{firstpage}--\pageref{lastpage}}
\maketitle

\begin{abstract}
Since its launch, \textit{TESS} has provided high cadence observations for objects across the sky. Although high cadence \textit{TESS} observations provide a unique possibility to study the rapid time evolution of numerous objects, artifacts in the data make it particularly challenging to use in studying transients. Furthermore, the broadband red filter of \textit{TESS}, makes calibrating it to physical flux units, or magnitudes, challenging. Here we present \tessreduce{} an open-source, and user-friendly Python package which is built to lower the barrier to entry for transient science with \textit{TESS}. In a few commands users can produce a reliable \textit{TESS} light curve, accounting for systematic biases that are present in other models (such as instrument drift and the varied \textit{TESS} background) and calculate a zeropoint to percent level precision. With this package anyone can use \textit{TESS} for science, such as studying rapid transients and constraining progenitors of supernovae.
\end{abstract}




\section{Introduction}

Built to discover and study exoplanets, the Transiting Exoplanet Survey Telescope (\textit{TESS}) has a unique mission profile of rapidly imaging large (24$^\circ\times$90$^\circ$) sectors of the sky \citep{Ricker2014}. During the prime mission, which ran from July 25$^{\rm th}$ 2018 to July 4$^{\rm th}$ 2020, \textit{TESS} recorded full frame images (FFIs) at a 30~minute cadence, however, this cadence was increased for the extended mission to 10~minutes. The combination of the missions wide filed of view, continuous coverage ($\sim$27 days per sector), high cadenced observations, and ability to perform precision photometry makes \textit{TESS} a unique instrument, not just for exoplanet studies, rather for all time varying phenomena. 

As we saw from the \textit{Kepler/K2} missions, 30~minute observations of transients can reveal new and unique features that are key to understanding progenitor systems of transient phenomena. Throughout the 8 year mission, \textit{Kepler/K2} observed numerous supernovae revealing features, in never-before seen time resolution. For example, in a study of 3 core collapse supernovae (CCSNe) observed by \textit{Kepler}, \citet{Garnavich2016} detected the first known shock breakout from a Type II SN (SN II) in KSN~2011d; this result was used to strongly constrain the progenitor star and SN properties, as well as theoretical explosion models. Because optical shock break out signatures only last for minutes to hours, high cadence observations from instruments such as \textit{Kepler} and \textit{TESS} are crucial to detecting and understanding this phenomenon. Similarly, \citet{Armstrong2021} presents the first full shock cooling curve of the CCSN AT~2017jgh. This unique data set again showed that it is crucial to capture the rise of the shock cooling curve at high cadence to effectively constrain models and obtain reliable progenitor properties. 

Precise high cadence observations are also crucial for understanding the progenitors of Type Ia Supernovae (SNe~Ia). Although these supernovae are known to come from the thermonuclear detonation of C/O white dwarfs in binary systems \citep[e.g.,][]{Hoyle1960, Woosley1986, Nomoto1984, Hoeflich1996, Hillebrandt2000, Bloom2012, Maoz2014}, and play a key role in understanding the evolution of the Universe as distance indicators \citep[e.g.,][]{Riess1998, Perlmutter1999, Betoule2014, Scolnic2018, Abbott2019}, the exact progenitor systems are unknown. This introduces a systematic uncertainty in the standardization of SN~Ia \citep[e.g.,][]{Foley2010,Polin2019}. Clues to the progenitor systems are expected to be found within a few days of explosion in the SN~Ia light curves. Nominally, SN~Ia are expected to rise according to the fireball model in which luminosity increases proportional to a power law with an index $\sim2$ \citep[$L\propto t^2$; e.g.,][]{Arnett1982,Riess1998,Olling2015,Nugent2011,Hayden2010,Miller2020,Wang2021}. However, different progenitor models predict deviations from the fireball model \citep[e.g.,][]{Piro2012, Firth2015}, which can sometimes be seen as excess flux at early times. \textit{Kepler/K2} saw such evidence of early excess flux in SN~2018oh \citep{Dimitriadis2018,Li2018,Shappee2018}, however, there was no evidence of excess flux in 5 other SN~Ia; in fact, SN~2018agk was shown to perfectly follow the fireball model \citep{Olling2015, Wang2021}.

In recent years, the number of rapid-evolving transients (RETs\footnote{Other acronyms used for RETs include fast-evolving luminous transients (FELTs) or fast blue optical transients (FBOTs).}), objects which rise to maximum brightness in $\lesssim$10 days and exponential decline in $\lesssim$30 days after rest-frame peak, have been increasing with samples such as those from Pan-STARRS1 \citep{Drout2014} and the Dark Energy Survey \citep{Pursiainen2018}, among others \citep[e.g.,][]{Arcavi2016b, Perley2020}. Transients like AT~2018cow \citep[e.g.,][]{Prentice2018, Ho2019, Margutti2019, Perley2019} and ZTF18abvkwla \citep[e.g.,][]{Coppejans2020, Ho2020} have been observed at multiple wavelengths. Additionally, KSN~2015K has been the only RET observed at extremely high cadence \citet{Rest2018}. While these unique events provided valuable insights into the nature of RETs, more high-cadence observations of the early light curves are key for understanding their explosion mechanism and progenitors.

Furthermore, distant supernovae can be gravitationally lensed and magnified by a foreground galaxy,  providing an opportunity to measure the time delay between light paths and constrain the Hubble constant. These gravitationally lensed SNe can be observed by \textit{TESS}. \citet{Holwerda2021} estimated the chances of observing SNe Ia and CCSNe being very low (0.6 and 1.3 per cent per year, respectively, and 2 and 4 percent at the ecliptic poles), but not null and highlighted the need for timely processing of \textit{TESS} imaging.

Alongside providing new insights into SNe, consistent high cadence observations also offer a new parameter space to search for new transients. \citet{Ridden-Harper2020} details a program to search the background pixels of the \textit{Kepler/K2} mission to identify previously unknown transients observed in high cadence, such as the WZ Sge type dwarf nova presented in \citet{Ridden-Harper2019}. Such a program could also be possible with \textit{TESS} which could provide us with a new window into a previously unexplored time domain.

These are all science cases in which \textit{TESS} could be a powerful tool for understanding transients. Some of these science cases are already being explored with the 4 years of data available. \citet{Fausnaugh2019} presented an analysis of 18 bright SN~Ia observed by \textit{TESS} and found no significant deviation from the fireball model at early times, however, they did observe slower rises than expected. A closer analysis of a SN~2018fhw, a bright SN~Ia observed by \textit{TESS}, also does not show any indication of early features in the light curve \citep{Vallely2019}. Similarly, a study of bright CCSNe observed by \textit{TESS} found no indication of a shock breakout at early times, suggesting the phenomena may be rare \citep{Vallely2021}. Furthermore, \textit{TESS} provided a comprehensive view of TDE ASASSN-19tb, showing its flux scales as $\propto t^2$ \citep{Holoien2019}. Similarly, \textit{TESS} observed an outburst of the repeating TDE ASASSN-14ko, showing it had a smooth powerlaw rise \citep{Payne2021}. Finally, \citet{Smith2021} presents the first \textit{TESS} light curve of a long GRB, in the detection of GRB 191016A.

Although \textit{TESS} has great potential for transient science, challenges with data reduction can present a significant hurdle. There are three components to \textit{TESS} data that must be addressed before the data can be used, which are:

\begin{itemize}
    \item Image alignment,
    \item Scattered light background, 
    \item Instrument calibration.
\end{itemize}

The first two components are key to extracting reliable \textit{TESS} light curves, while the final component is crucial for comparisons to supplementary data and model fitting. Currently, there are several successful reduction pipelines for \textit{TESS} FFIs (such as \textsc{eleanor};\citealt{Feinstein2019}), difference imaging pipelines (such as the DIA pipeline; \citealt{Oelkers2018}), the CDIPs pipeline; \citealt{Bouma2019}), those that make use of the ISIS package \citep{Alard1998,Alard2000} (such as \citealt{Vallely2019}, \citealt{Fausnaugh2019}, \citealt{Vallely2021}), and the LINEAR pipeline \citep{Woods2021}. However, these methods have limitations, particularly in addressing the highly variable background scattered light, and flux calibration. We examine these limitations in \autoref{sec:background}, and \autoref{sec:calibration}. While \textsc{eleanor} is pip installable\footnote{\url{https://adina.feinste.in/eleanor/}}, the other methods require local installs of complex reduction pipelines, making \textit{TESS} data reduction challenging. With \tessreduce{}, we provide another open-source user-friendly python package that addresses all three challenges in dealing with \textit{TESS} data, without the need for high profile difference imaging pipelines.

In this paper, we present the \tessreduce{} pipeline for \textit{TESS} data analysis, alongside light curves of selected transients. As seen in \autoref{fig:reduction_example} even in cases of extreme background \tessreduce{} can recover transient signals. We highlight the data products used by \tessreduce{} in \autoref{sec:data}; the background determination method in \autoref{sec:background}; image alignment in \autoref{sec:align}; absolute photometric calibration of \textit{TESS} photometry in \autoref{sec:calibration}; and finally, transient light curves reduced by our method/code are presented in \autoref{sec:light curves}.

Our code is open source and can be accessed on Github\footnote{\url{https://github.com/CheerfulUser/TESSreduce}} or pip installed. The \tessreduce{} docs can be found online\footnote{\url{https://tessreduce.readthedocs.io/en/latest/}}. \tessreduce{} has already been used to enable analysis of the gamma ray burst GRB~191016A \citep{Smith2021}, and the peculiar comet C/2014 UN271 \citep[Bernardinelli-Bernstein;][]{Ridden-Harper2021}.

\section{Data} \label{sec:data}
To produce reliable light curves from \textit{TESS} data, \tessreduce{} leverages data from across numerous instruments. This data is used in background determination and instrument photometric calibration. Here we discuss all data products used in \tessreduce{}.

\subsection{TESS}

\begin{figure*}
    \centering
    \includegraphics[width=\textwidth]{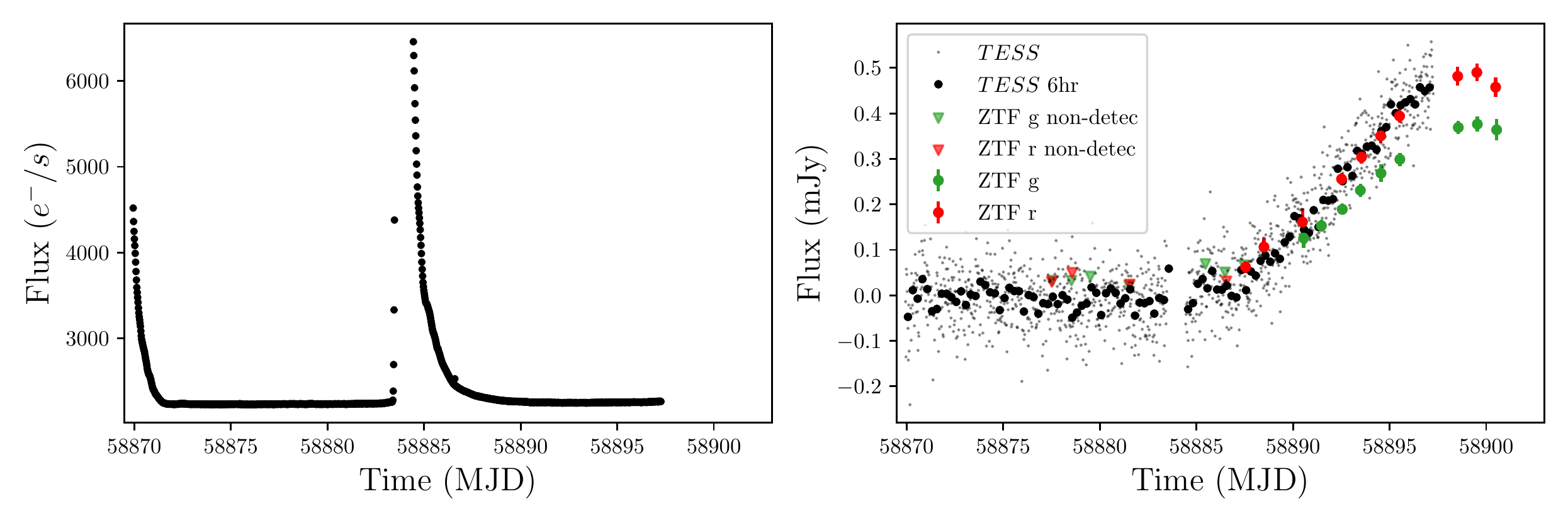}
    \caption{\textit{TESS} light curves of the rise of SN~2020cdj. \textbf{Left:} Raw \textit{TESS} light curve, this is dominated by background scattered light which cause the spikes in the light curve around 58870 and 58885 MJD. As a comparison the scatter light background peaks with a brightness of over 6000~counts, while SN~2020cdj peaks at $\sim20$~counts. \textbf{Right:} Reduced and flux calibrated \textit{TESS} light curve created by \tessreduce{}, alongside public ZTF $g$ and $r$ observations. With the correct reduction, \textit{TESS} is capable of reaching sensitivities comparable to state-of-the-art ground based survey telescopes, with higher cadence.}
    \label{fig:reduction_example}
\end{figure*}

\begin{figure}
    \centering
    \includegraphics[width=\columnwidth]{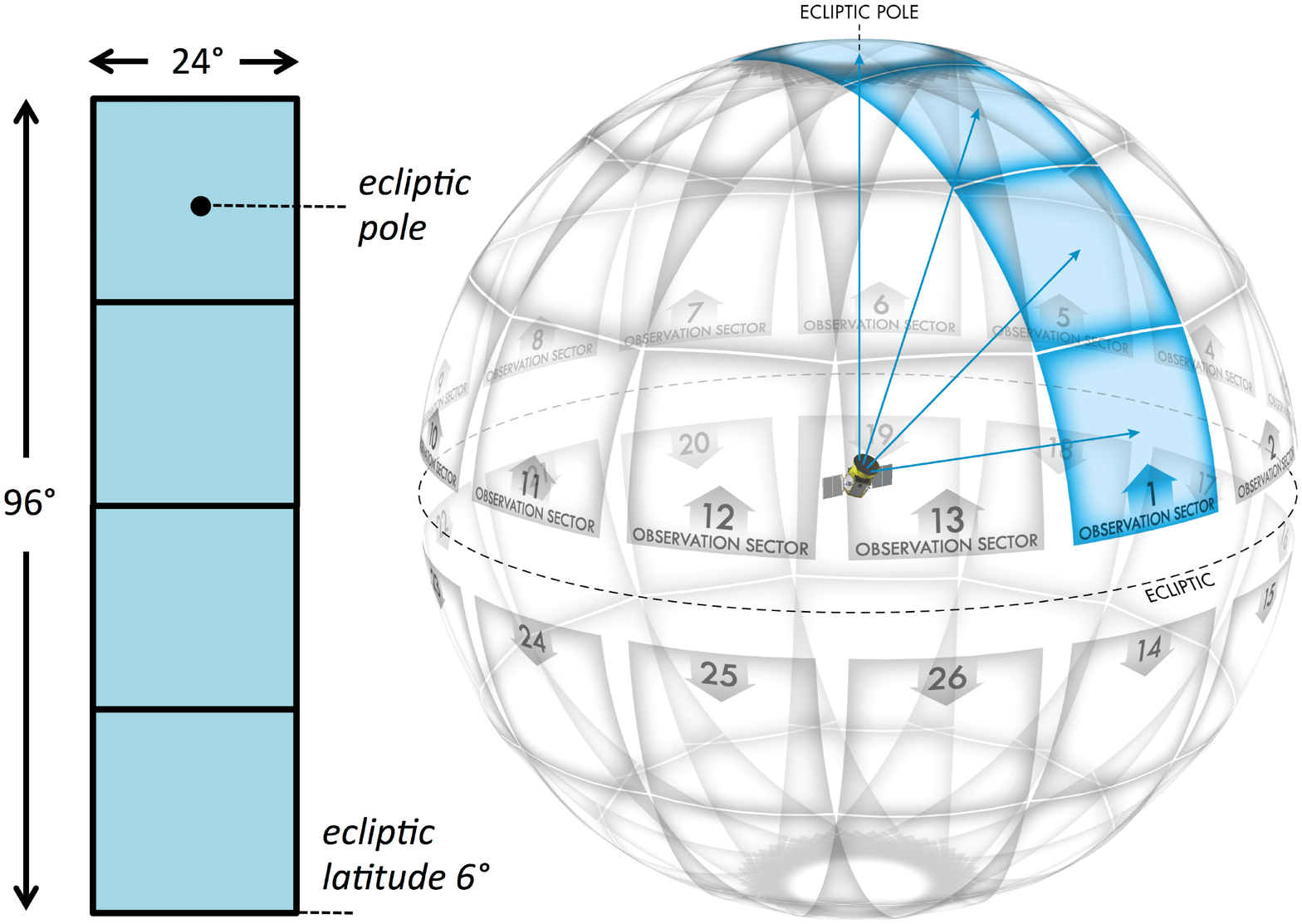}
    \caption{Schematic of \textit{TESS} sectors adapted from \citet{Ricker2014} Fig. 7.}
    \label{fig:tess_sectors}
\end{figure}

\textit{TESS} data is characterized by low spatial resolution (21\arcsec{} per pixel), but high temporal resolution. The observing campaigns, known as sectors, last for $\sim28$~days, with a mid-sector data gap at $\sim14$~days. As seen in \autoref{fig:tess_sectors}, regions toward the poles can have multiple sectors of overlap due to the strip-like nature of the \textit{TESS} field-of-view, extending the \textit{TESS} coverage to a maximum of a year at the poles. 

The three primary data products are the Full Frame Images (FFIs), Target Pixel Files (TPFs), which cover small regions of the detector at higher cadence, and SPOC pipeline lightcurves. The data formats for the FFIs and TPFs are also different, where the FFIs are individual images, while the TPFs are data cubes made from time series of images. In the primary mission, which consists of data from sectors 1--26, FFIs were recorded at a 30~minute cadence, while the targeted TPFs had a 2~minute cadence. For the extended mission, which will cover sectors 27--onwards, the cadence was increased to 10~minutes for FFIs and 20s for TPFs.

\tessreduce{} was built using the \textsc{lightkurve} package as a foundation, therefore it utilizes the TPF data format \citep{lightkurve}. Although FFIs are not directly compatible with \tessreduce{}, TPF cutouts can readily be created from the FFIs using the \texttt{TESScut} tool \citep{TESScut}. In general, we use comparably large $90\times 90$~pixel cutouts for \tessreduce{}, to generate reliable determinations of the complex background, since point sources are nominally $3\times3$~pixels in size. 

\subsection{Pan-STARRS (PS1)}
To supplement the \textit{TESS} data, we make extensive use of the (PS1) $grizy$ photometry from the 1.8~m Pan-STARRS 1 telescope (PS1) \citep{Tonry2012}. PS1 covers the Northern sky to a declination of $-30^\circ$ and limiting magnitudes of 23.3, 23.2, 23.1, 22.3, 21.4~mag, for the $grizy$ filters, respectively. Sources brighter than $\sim14$~mag reach saturation in PS1 so are not included in the source catalogs, such as the DR1 catalog we use here \citep{Chambers2016}. 

Alongside source masking, we use PS1 data for photometric calibration of \textit{TESS}. The total PS1 system is superbly calibrated and consistent, with $<1\%$ variability across the detector \citep{Schlafly2012}. Furthermore, to achieve precise cosmological measurements, the PS1 photometric system has been well calibrated with numerous probes. As described in \citet{Scolnic2015}, Supercal calibration achieved a relative calibration to better than $\sim5$~mmag with Calspec sources. \citet{Narayan2019} extends the Calspec catalog with faint DA white dwarfs which have hydrogen-dominated atmospheres, and also finds the PS1 system is consistent to mmag levels of precision. Due to the well calibrated nature of PS1 photometry and wide wavelength coverage, we primarily use PS1 photometry for photometric calibration. 

We obtain the PS1 data from Vizier table II/349/ps1 with \textsc{astroquery} \citep{Astroquery}. This high resolution and multi-band data is crucial for both generating a source mask, and calibrating the \textit{TESS} photometry. 

\subsection{SkyMapper}
To complement the PS1 photometry, and fill in the Southern sky, we use $griz$ photometry from the 1.3~m SkyMapper telescope \citep{Keller2007}. We use the DR1 catalog presented by \citet{Wolf2018}, which is complete to $\sim18^{\rm th}$~mag, with a saturation limit of $\sim9^{\rm th}$~mag in all filters. Even in this early data release, SkyMapper provides an excellent complement to PS1 where photometry between the two systems is consistent to within an RMS scatter of 2\%. 

As with the PS1 DR1 source catalog, we access the SkyMapper DR1 catalog with \textsc{astroquery} and Vizier table II/358/smss.

\subsection{Gaia}
We rely on Gaia data to supplement the PS1 and SkyMapper data, particularly for saturated sources. Gaia has produced an all sky catalog of sources with a limiting magnitude of 22 and a saturation magnitude of $\sim3$ \citep{GAIA,GAIA_DR2,GAIA_catalog}. We access the Gaia DR2 catalog I/345/gaia2 through Vizier, with \textsc{astroquery}. Although the Gaia \textit{G} filter magnitudes can be a close approximation to \textit{TESS} magnitudes, as shown in \citet{Stassun2019}, we elect to only use Gaia data in the creation of source masks, since we can achieve more reliable \textit{TESS} magnitudes from PS1 and SkyMapper photometry. 

\subsection{Calspec}
To construct a photometric mapping from PS1 magnitudes to \textit{TESS} magnitudes and an extinction coefficient for \textit{TESS}, we rely on the well calibrated spectra within the Calspec library\footnote{\url{https://www.stsci.edu/hst/instrumentation/reference-data-for-calibration-and-tools/astronomical-catalogs/calspec}} \citep{Bohlin2014,Bohlin2020}. In this application of the catalog, we restrict ourselves to the use of only the gold-standard Calspec sources with Hubble STIS coverage, and drop the series of sources associated with NGC~6681 as we find that they significantly deviate from the canonical stellar locus. 

\subsection{ZTF photometry}
For transients, we use the public ZTF photometry to compare against the calibrated \textit{TESS} light curves. Although we do not directly use these light curves in the \tessreduce{} reduction, they are a useful tool for verifying the integrity of the \tessreduce{} light curve and identifying host flux. An example of this comparison can be seen in Fig~\ref{fig:reduction_example}. To gather the ZTF light curves we use the Open Supernova Catalog \citep[OSC,][]{Guillochon2017} application programming interface (API) to identify the ZTF-specific transient name, which can then be used jointly with the Automatic Learning for the Rapid Classification of Events \citep[ALeRCE,][]{Forster2021} API to download public ZTF $g$ and $r$ band photometry.

\section{Background determination} \label{sec:background}
The greatest challenge with \textit{TESS} data is in accurately determining and removing the background caused by scattered Earth and Moon light. Due to the close 28~day orbit \textit{TESS} has between the Earth and Moon, it is prone to suffering from significant levels of background from scattered light. In \autoref{fig:reduction_example}, we show the raw lightcurve for SN~2020cdj (left) which is dominated by the scattered light background, peaking at over 6000~counts, in contrast SN~2020cdj peaks at $\sim20$~counts. Such scenarios where the background signal is $10-100$ times brighter than the target are common. This background is highly variable, both spatially and temporally, which is usually repeated for the two 14~day sector segments. To compound the background challenges, the low resolution images lead to source blending, often leaving few true sky pixels from which the background can be determined. Furthermore, the background is comprised of two distinct aspects, a smooth and continuous background, and a discrete vertical background caused by electrical straps behind the CCDs. 

Different image reduction pipelines account for the background with different methods. For example both the DIA and CDIPS pipelines rely on median smoothing the background with a $32\times32$~pixel and a $48\times48$~pixel grid, respectively. This method will biased the background to higher values, especially in crowded fields. Furthermore, the background structure varies on smaller distance scales, meaning these large bins will miss background structure. As the two image-reduction pipelines are described, they also do not account for the strap artifacts, which will lead to poor background subtractions for sources that fall in the strap columns. The LINEAR pipeline, however, does not list a background determination and subtraction method. 

The ISIS-based difference imaging pipelines such as those used in \citet{Vallely2019} and \citet{Fausnaugh2019} determine the background \textit{after} the reference image is subtracted. After subtraction, the background is then calculated in a similar manner to the DIA and CDIPS pipelines by calculating the median background from $30\times30$~pixel regions. While this method significantly reduces the risk of sources biasing the background, it will still be biased by the cores of bright objects that do not fully subtract out. Likewise, the enhanced strap background is not accounted for in these pipelines. \citet{Vallely2021} improves upon the background subtraction by decomposing the 2D background into 1D strips. First they calculate and subtract the vertical component of the 1D background by median filtering individual columns of the subtracted image over 30 rows. The column background is then subtracted from the difference image, limiting the effects of the strap background, before the same process is applied to the rows. Although this method can produce a close approximation of the complex \textit{TESS} background, it has some limitations. Mainly, the straps in this method are treated as an additive component, however, they are a multiplicative component, arising from the straps scattering light back into the CCD, effectively enhancing the quantum efficiency (QE) for strap columns \citep{TESShandbook}. Furthermore, like with the other methods, imperfections in the difference image will bias the background, particularly around bright and variable sources. 

With \tessreduce{}, we address the shortcomings in background determination of other difference imaging pipelines by masking sources and addressing the discrete strap background according to the physical origin.

\subsection{Source/Image masking}
A key aspect in determining the \textit{TESS} background is to first identify all sources and key features within a \textit{TESS} image. To create an accurate source mask we use both the PS1 DR1 and Gaia DR2 source catalogs to identify all sources within an image, to a depth of 19th mag in PS1 $i$. To generate the preliminary source mask, we separate sources into magnitude bins, which have individual mask aperture sizes. For sources with $i>14$ we use simple square apertures, however, for sources with $i<14$, we use the Gaia \textit{G} band and construct a 2-component mask, comprised of a central square mask and a secondary cross mask, to cover diffraction/bleeding columns. With these masks we identify and isolate all known sources.

Although the preliminary mask covers most known sources, extended sources, such as nearby galaxies, often are not fully contained within the preliminary mask. To correct this, \tessreduce{} constructs a reference image by selecting an image where the background is low and well behaved/uniform. We took images with total summed flux in the lowest 5$^{\rm th}$ percentile to be the low background images.  We then apply the preliminary source mask and cut all pixels brighter than $2\sigma$ above the median. This strict limit is chosen to minimize the number of pixels contaminated by sources in the background sample. All clipped pixels are added in to the source mask. This captures larger diffuse sources in the mask. 

With the source mask, we now construct a bit mask identifying all features in an image. We assign distinct values to unsaturated sources, saturated sources, and strap columns, including four adjacent columns on either side. This mask allows us to readily distinguish between background, and source pixels, alongside columns that are influenced by straps.

In crowded fields, the default choices we make to construct the mask can leave few pixels remaining to estimate the background. For all masks we check the mask coverage fraction by taking the ratio of masked pixels, to total pixels. If the total source mask covers $>90\%$ of the image, we weaken the catalog mask requirements to a limiting magnitude of $i=16$ and reduce the default mask aperture by half. Although this compromise makes enough pixels available to determine the background, it will now includes sources, thus reducing the background quality. In the general reduction process, the effect of unmasked objects is reduced as the final background is calculated for difference images, where most of the flux is removed for non-variable sources.

\subsection{Smooth background determination}
\begin{figure*}
    \centering
    \includegraphics[width=\textwidth]{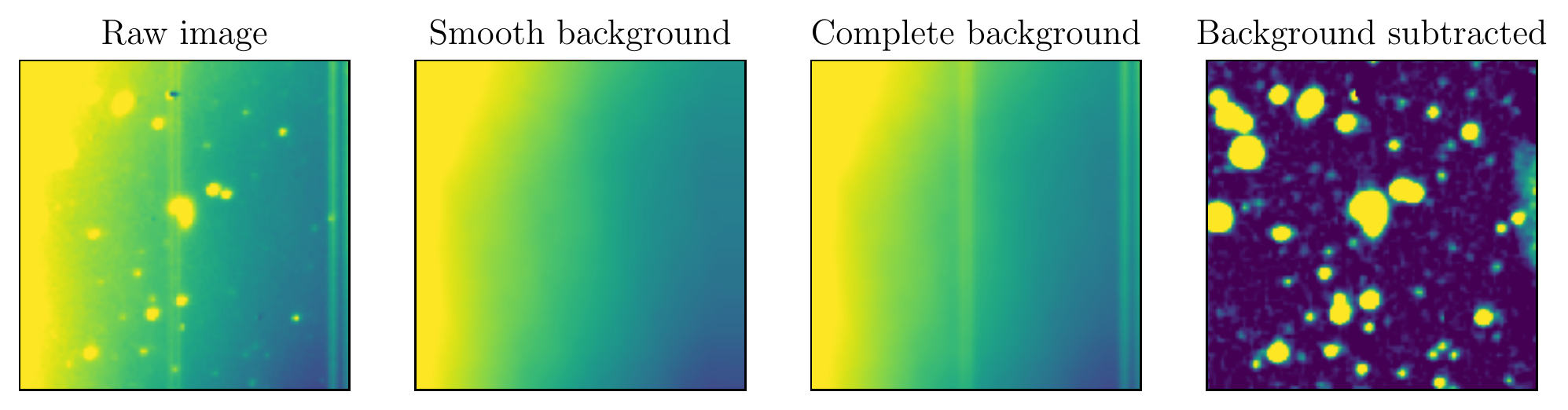}
    \caption{Example background subtraction for a $90\times 90$~pixel cutout of the \textit{TESS} FFI centered on SN~2020fqv. This region features multiple strap features, highly variable background, and  numerous complex sources. \textbf{Left:} Raw \textit{TESS} image at a time of moderate background flux. \textbf{Middle left:} Smooth background component determined by \tessreduce{} by interpolating over masked sources and strap columns. \textbf{Middle right:} Smooth background, plus the effective quantum efficiency amplification produced by electrical straps running behind the CCD. \textbf{Right:} Background subtracted \textit{TESS} image, some background residuals still remain on the right side of the image since those columns coincide with straps and were masked when calculating the smooth background. The colour scale for the first 3 images spans 8000 to 9000 counts, while the colour scale for the subtracted image spans 0 to 50 counts.}
    \label{fig:background}
\end{figure*}

Determining a reliable smooth background for the \textit{TESS} images is key in \tessreduce{}, as it is also used in determining the discrete strap backgrounds. Determination of the smooth background follows on simply from the masking process. For each image we mask out all pixels with non-zero bits in the bit mask, which includes the strap columns. We then linearly interpolate the background over the missing pixels with the \texttt{scipy.griddata} algorithm. Since the linear interpolation method of \texttt{griddata} is unable to extrapolate, we fill any edge regions that require extrapolation using the ``nearest'' method. Finally, we smooth the background using a 2D Gaussian filter with $\sigma=1$~pixel. An example of this background is shown in \autoref{fig:background}.

\subsection{Strap background determination}
As mentioned previously, the discrete background is caused by electrical straps behind the CCD scattering light back into the CCD. The scattering is colour dependent and effectively acts to enhance the QE of columns backed by straps and their neighbors. We isolate the straps through the bit mask and omit them from the smooth background determination. To correct for the straps, we calculate their QE enhancement by comparing the strap background to the calculated smooth background. 

To calculate the apparent QE for strap columns, we use the bitmask to mask all sources and columns not associated with straps. We then divide the masked image by the previously calculated smooth background for each image. Although the enhancement is colour dependent and exhibits variation along an entire FFI strap column, for small cutouts, such as those used in \tessreduce{}, there is little variation, so we take the enhanced QE for a column to be the median of all sky pixels. Generally, the enhancement is small at around 2\%, peaking at $\sim6\%$ in regions suffering from extreme levels of scattered light. Finally, we multiply the smooth background by the calculated QE to obtain a background that encapsulates both the smooth and discrete components which is then subtracted from the raw images.


\section{Image alignment} \label{sec:align}
Although \textit{TESS} is largely stable, due to the large pixel size, small changes in pointing throughout a sector, caused either by instrument drift or relativistic aberration effects, can lead to significant trends in a light curve. These effects are particularly present for transients in crowded fields, or within an extended nearby galaxy. To calculate the relative shifts of each image to a reference we use the \texttt{DAOStarFinder} function in the \texttt{photutils} package \citep{photutils2020}. This method locates all centroids in an image and provides their fractional pixel position. We then calculate the $x,y$ pixel offset for each of the sources in every image, relative to the reference. Although the changes in apparent \textit{TESS} pointing will be a combination of spacecraft roll around the boresight, jitter, and relativistic effects, on small scales such as used in \tessreduce{}, it is sufficient to treat it as a simple linear shift. 

Although we now have offsets for each image, they contain significant scatter. Therefore, aligning images with these offsets will introduce significant scatter to the light curves. To limit the scatter introduced by aligning images, we smooth the shifts with a Savitzky–Golay filter with a window size of 51 images and a third order polynomial. We use these smoothed $x,y$ pixel offsets to align each image to the reference using the \texttt{shift} algorithm from the \texttt{scipy ndimage shift} package.

Ideally, we would shift the images before calculating and subtracting the background, however, we found that high background levels can skew the offsets to incorrect values. To mitigate this, we take the iterative approach of first determine and subtract the background from the raw images, calculate the image offsets, and then restart the reduction, thus time aligning the images according to the previous offsets before calculating the background. Although this iterative approach increases the computation time for each reduction, it ensures we have reliable shifts that are not influenced by the variably \textit{TESS} background.

\begin{figure}
    \centering
    \includegraphics[width=\columnwidth]{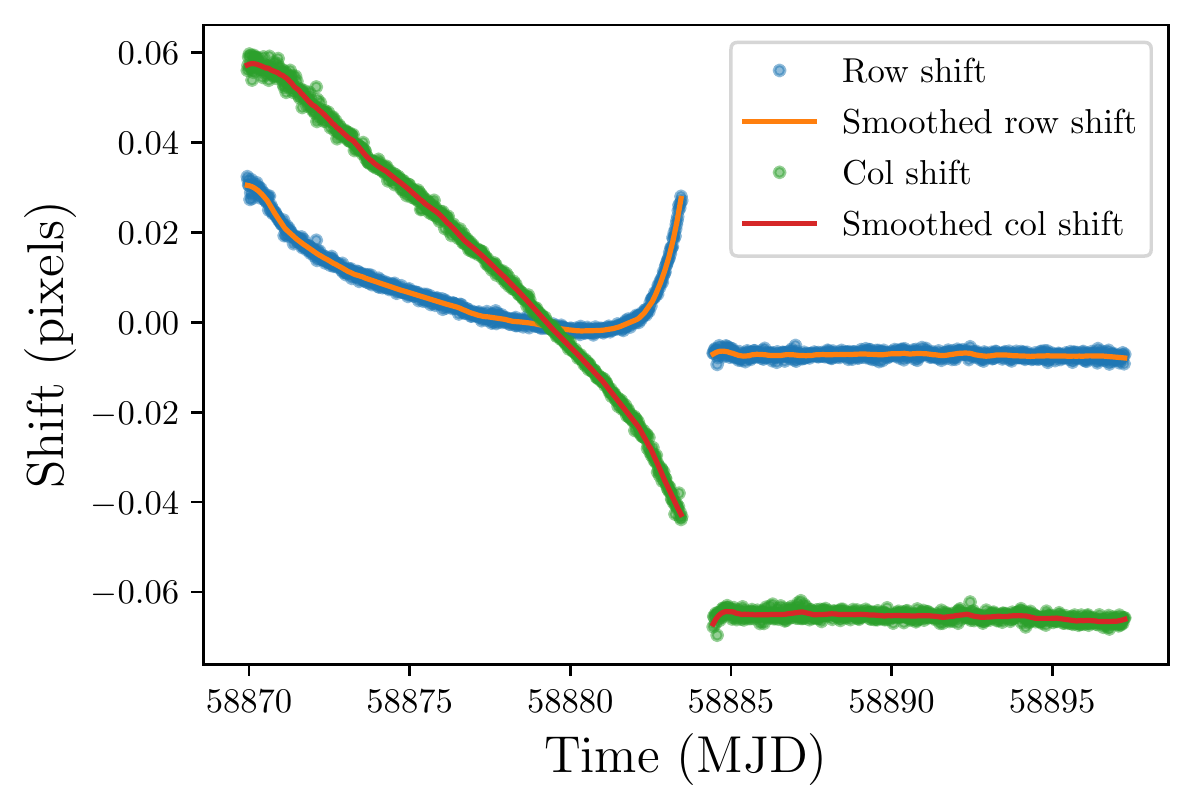}
    \caption{\textit{TESS} pointing shifts for SN~2020cdj in sector 21, camera 4, CCD 3. While these shifts are small, only ever a fraction of a pixel, they have a significant effect on light curves, substantially changing their shapes. These shifts were found using the \texttt{photutils} \texttt{DAOStarFinder} algorithm and smoothed with a Savitzky–Golay filter.}
    \label{fig:align}
\end{figure}

Generally, we find that the \textit{TESS} alignment shifts by only a few percent across each sector. As seen in \autoref{fig:align}, the the offset pattern is quite variable and can behave differently for the two orbits in a sector.

\subsection{Kernel matching}
An alternative to shifting images, is to align them through convolving with Delta kernels. This approach is used by the ISIS based pipelines \citep[e.g.][]{Fausnaugh2019}, and accounts for both relative shifts between images, and any changes in the point spread function (PSF). For ground based telescopes, variable atmospheric conditions lead to large differences in the point spread function (PSF) between subsequent observations that must be corrected through the process of kernel matching. For \textit{TESS}, there is minimal change in the instrumental PSF and sampled pixel response function (PRF). While accounting for the subtle changes in the PRF in \textit{TESS} data would also improve light curve quality, accurately determining differences in PRFs between \textit{TESS} images with the small cutouts used by \tessreduce{} is challenging. In \autoref{sec:shift_vs_delta}, we compare the image alignment method used in \tessreduce{} to a locally determined Delta kernel shift and find that the image alignment method produces more reliable results.

\section{Photometric calibration} \label{sec:calibration}
While all \textit{TESS} FFIs available on the Mikulski Archive for Space Telescopes (MAST\footnote{\url{https://archive.stsci.edu/}}) are calibrated, they are not calibrated to astrophysical flux units, rather, to counts in terms of $e^-/s$ (electrons per second). To make comparisons of \textit{TESS} light curves to other data or models, we must have a reliable way of calibrating the \textit{TESS} photometry such that we have a zeropoint that converts \textit{TESS} counts to astrophysical units. 

Other pipelines, such as the LINEAR-\textit{TESS} pipeline, calibrate a \textit{TESS} photometry to Gaia photometry \citep{Woods2021}. Using non-variable stellar sources, they match \textit{TESS} instrumental magnitudes to Gaia \textit{G} magnitudes to calculate a zeropoint of $21$. While the \textit{TESS} and Gaia \textit{G} filters share some similarities, \textit{G} extends to much shorter wavelengths than \textit{TESS}, making it unreliable to directly compare the two systems (see \autoref{fig:gaia_tess} for a comparison). 

Alternatively, the DIA pipeline calculates zeropoints by comparing instrumental magnitudes to \textit{TESS} magnitudes for all stars presented in the \textit{TESS} Input Catalog (TIC). This method derives a zeropoint of 4.825, which is substantially lower than zeropoints derived via other methods. This difference is likely due to re-scaling in the reduction process.

Through tests in the commissioning phase, \citet{TESShandbook} derives an average zeropoint across the detectors of $z_p=20.44\pm0.5$. While the average zeropoint is a good estimate, it is highly uncertain, so for reliable flux conversions we must calculate the zeropoint to higher precision. Furthermore, the zeropoint will also be dependent on the choice of reduction, so, for consistent flux-calibrated \textit{TESS} light curves, we have developed a method for \tessreduce{} to perform in-situ calibration from field stars.

While broadband filters such as that of \textit{TESS} can be challenging to calibrate, we utilize the calibration method developed in \citet{Ridden-Harper2021b} which calibrates \textit{Kepler/K2} broadband photometry to a sub-percent level. This method uses the Calspec spectral library alongside the well calibrated PS1 DR1 photometry to predict magnitudes in broadband filters. We adapt this process to calibrate \textit{TESS} photometry through the process described in the following section.

\subsection{Mapping PS1 \& SkyMapper to \textit{TESS} magnitudes}
\begin{figure}
    \centering
    \includegraphics[width=\columnwidth]{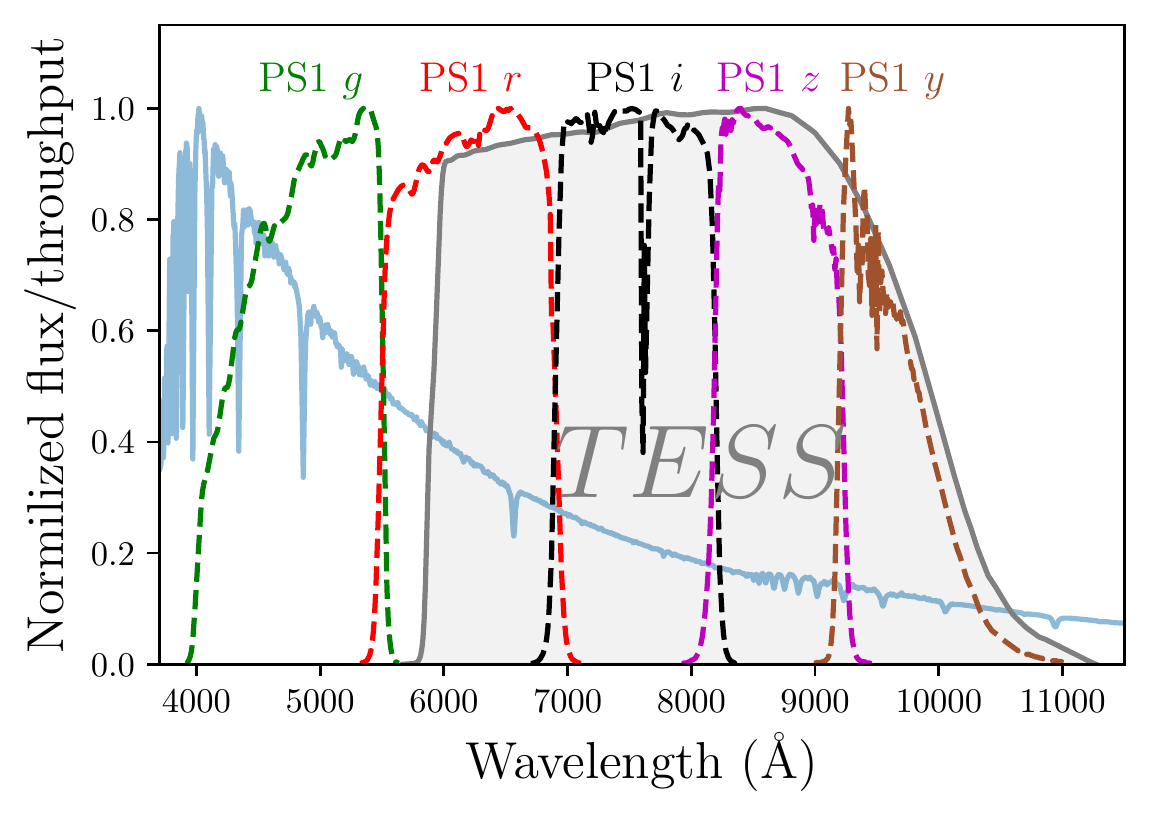}
    \includegraphics[width=\columnwidth]{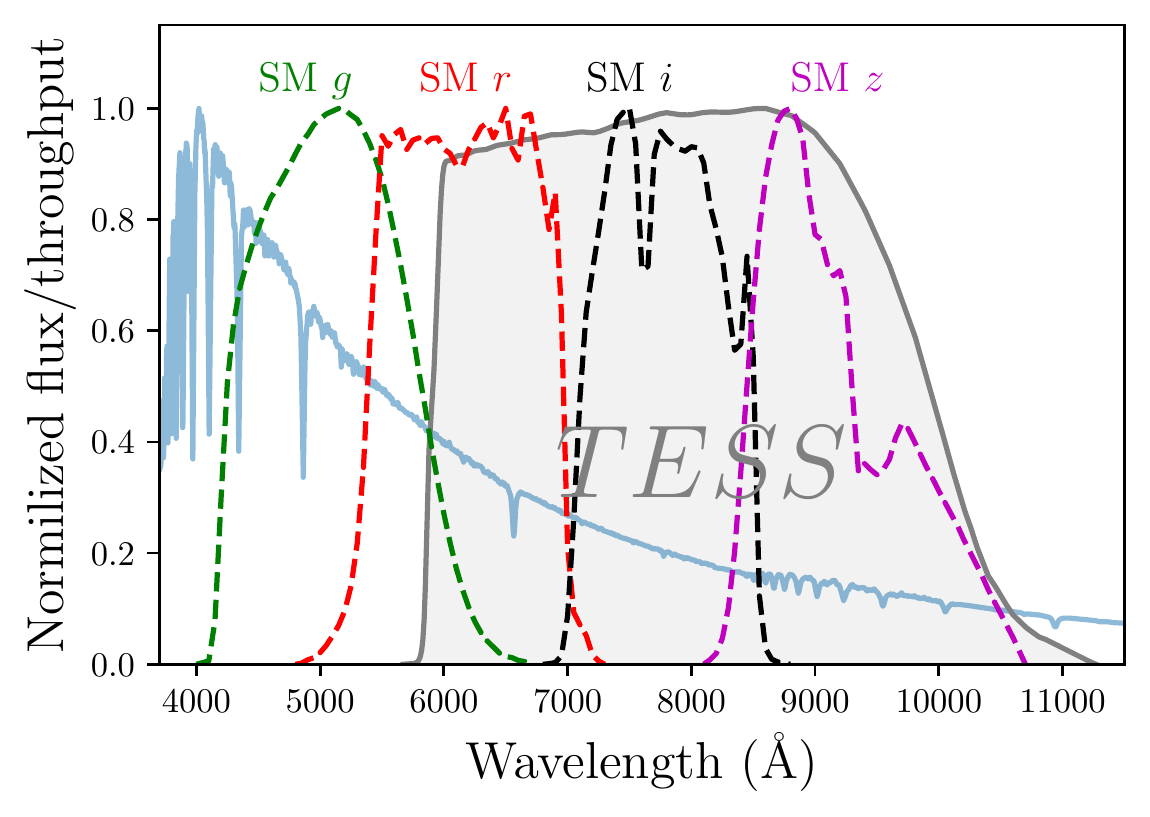}
    \caption{Comparisons of the \textit{TESS} filter overlaid on the Calspec spectra for HD~2811 alongside the PS1 $grizy$ (\textbf{top}) and the SkyMapper $griz$  (\textbf{bottom}) filters. It is clear that the PS1 $rizy$ and SM $riz$ completely overlap \textit{TESS}, so we can approximate the \textit{TESS} filter with a combination of PS1 or SM filters.}
    \label{fig:calspec}
\end{figure}

Following \citet{Ridden-Harper2021b}, we assume that the \textit{TESS} filter can be represented as a linear combination of PS1 or SkyMapper filters in flux space with a linear color correction term in magnitude space. As seen in \autoref{fig:calspec}, the \textit{TESS} filter extends from $5802.57$ to $11171.45$~\AA, which covers the PS1 $rizy$ filters, and, for the color correction, we take the difference between $g$ and $i$. The large wavelength difference between $g$ and $i$, gives us a good baseline for identifying any colour terms.

To determine the contributions from each of the PS1 filters, we use synthetic photometry on the Calspec spectral library. With the filter definitions of the PS1 and \textit{TESS} filters, obtained from the Spanish Virtual Observatory\footnote{\url{http://svo2.cab.inta-csic.es/theory/fps/}} \citep[SVO;][]{SVO2012,SVO2020}, we calculate the synthetic magnitude of each Calspec source in all filters. We minimize the magnitude difference between the composite \textit{TESS} magnitude produced by the PS1 filters and the \textit{TESS} magnitude across all sources. This gives us the following relation in flux space for a zeropoint of 25, to match the PS1 image zeropoints:

\begin{eqnarray}
f_{tess}^{PS1} \approx (0.256f_r^{PS1} + 0.276f_i^{PS1} + 0.358f_z^{PS1} + 0.112f_y^{PS1}) \nonumber\\ \times\left(\frac{f_g^{PS1}}{f_i^{PS1}}\right)^{0.001}.
\end{eqnarray}
Using the same method we can construct an approximation of the \textit{TESS} filter with the SkyMapper $griz$ filters to get the following:
\begin{eqnarray}
f_{tess}^{SM} \approx (0.258f_r^{SM} + 0.353f_i^{SM} + 0.394f_z^{SM})\left(\frac{f_g^{SM}}{f_i^{SM}}\right)^{-0.002}.
\end{eqnarray}

\begin{figure}
    \centering
    \includegraphics[width=\columnwidth]{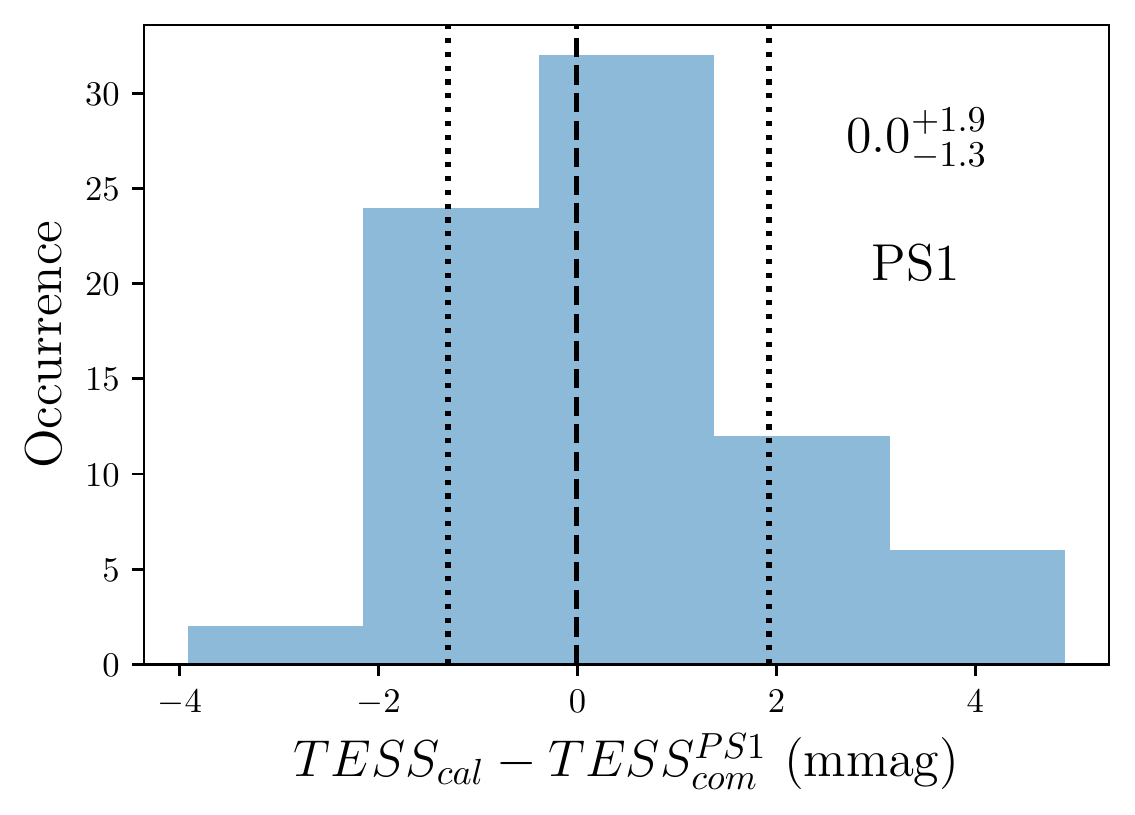}
    \includegraphics[width=\columnwidth]{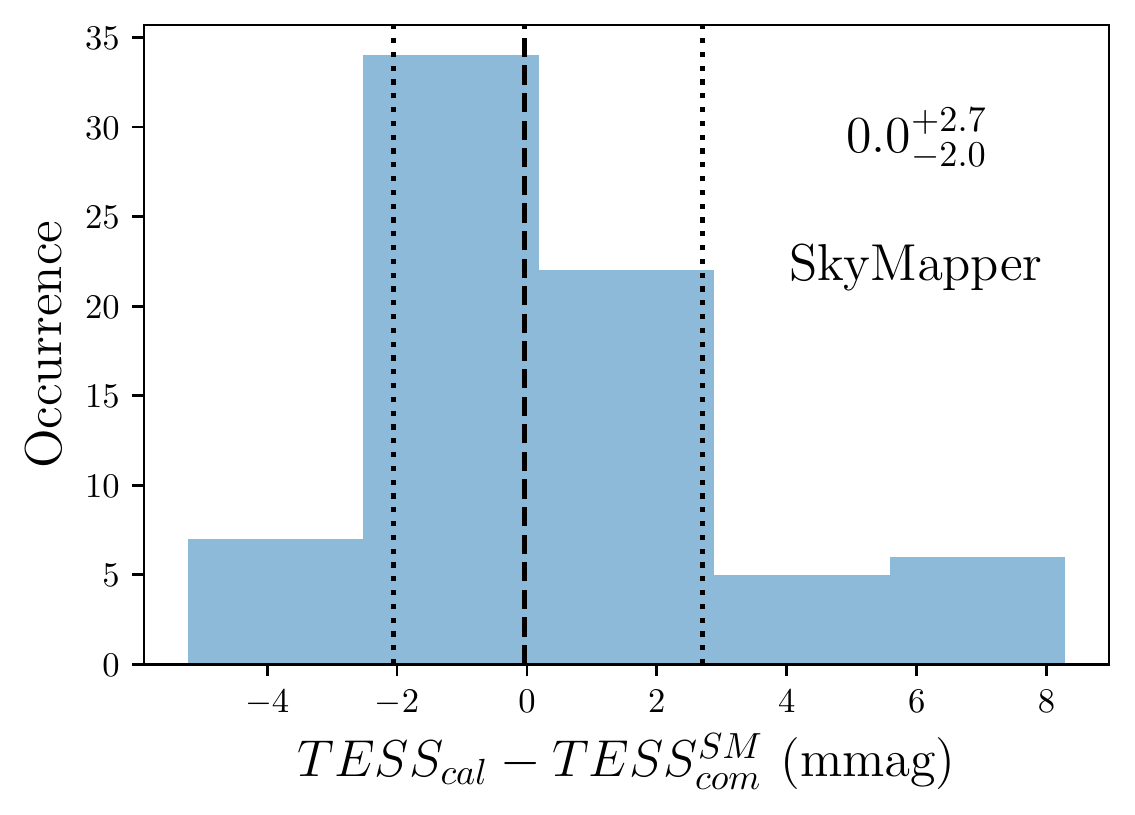}
    \caption{Residuals of the synthetic and composite \textit{TESS} magnitudes created from Calspec sources, shown with the median and errors represented by the $80^{th}$ and $16^{th}$ percentiles. Both the composite constructions of the \textit{TESS} filter with PS1 $grizy$ (\textbf{top}) and SkyMapper $griz$ (\textbf{bottom}) produce reliable approximations for stellar sources to the mmag level.}
    \label{fig:syn_cal}
\end{figure}

Both of these relations provide an excellent approximation for the \textit{TESS} filter. As seen in \autoref{fig:syn_cal}, the residuals are consistent with zero at the mmag level. These relations are weakly dependent on colour, so, although it was constructed using stellar sources between $-0.5<(g-r)<0.8$, it is reliable for other stellar sources, though the error will grow. 

With these relations we can confidently predict the expected \textit{TESS} magnitude of a well behaved stellar source using PS1 \& SkyMapper DR1 photometry. This forms the basis of our calibration method.

\subsection{Dust extinction} \label{sec:dust}
\begin{figure*}
    \centering
    \includegraphics[width=\textwidth]{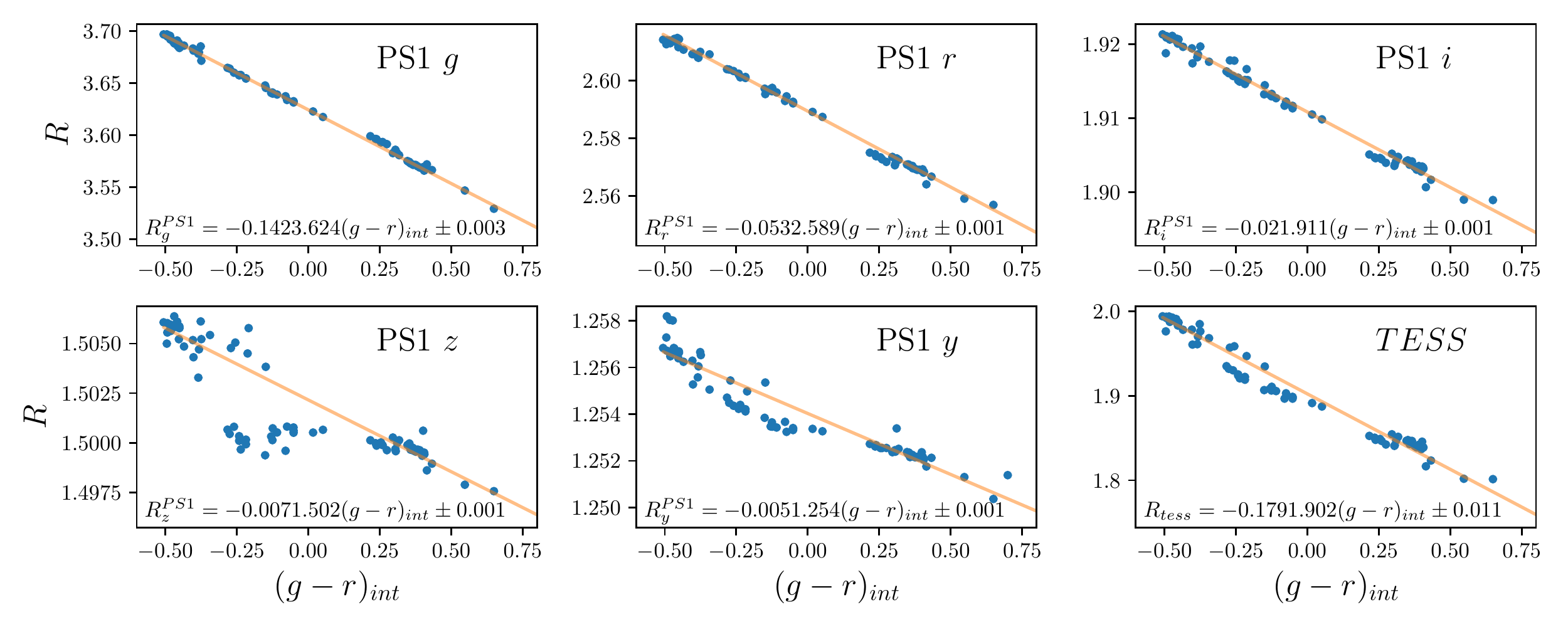}
    \caption{Extinction vector coefficients, $R$, for the PS1 $grizy$ and $TESS$ filters. For larger color ranges the relation becomes non-linear as seen in \citet{Tonry2012}, however, the color range of $-0.5<g-r<0.8$ is sufficient for the calibration used in \tessreduce{}. The equations for each fit alongside the 1$\sigma$ error.}
    \label{fig:R_fit}
\end{figure*}

For accurate photometric calibration, we must also account for how dust extinction impacts \textit{TESS} photometry. While dust extinction effects an entire spectrum, the cumulative effect on photometry is that the observed magnitude, $m_{obs}$, will be changed according to:
\begin{eqnarray}
m_{obs}=m_{int}+E(B-V)R_x, \label{eqn:extinction}
\end{eqnarray}
where $m_{int}$ is the intrinsic magnitude, $E(B-V)$ is the ($B$-$V$) colour excess, and $R_x$ total-to-selective extinction in the $x$ filter. While $R_x$ is known for the PS1 filters, we must define it for \textit{TESS} and the SkyMapper $griz$ filters. To do this, we follow the derivation used in \citet{Tonry2012} and adapted in \citet{Ridden-Harper2021b} to use synthetic photometry to determine $R_x$.

We derive $R_x$ by comparing synthetic magnitudes of normal and reddened Calspec spectra. We redden the Calspec spectra by applying an extinction of $E(B-V)=0.01$, with $R_V=3.1$ using the \citet{Fitzpatrick1999} dust law through the \texttt{extinction} package \citep{Barbary2016}. Next, we calculate the synthetic magnitude of the reddened Calspec sources in the \textit{TESS}, PS1 $grizy$, and SM $griz$ filters. The difference between the normal and reddened magnitudes gives us the extinction vector coefficient simply by rearranging \autoref{eqn:extinction}. As seen in \autoref{fig:R_fit} (and \autoref{fig:R_skymapper}), $R_x$ varies with colour, but is well described by a straight line. For each filter we fit $R_x$ by minimizing the $\chi^2$ with the \texttt{scipy} \texttt{minimize} algorithm. Here we show the \textit{TESS} $R_x$ alongside the SM $griz$ vectors and the PS1 $grizy$ vectors:
\begin{eqnarray} 
R^{PS1}_g=3.624-0.142(g-r)_{\rm int},\nonumber\\
R^{PS1}_r=2.589-0.053(g-r)_{\rm int},\nonumber \\
R^{PS1}_i=1.911-0.02(g-r)_{\rm int},\nonumber\\
R^{PS1}_z=1.502-0.007(g-r)_{\rm int},\nonumber\\
R^{PS1}_y=1.254-0.005(g-r)_{\rm int},\nonumber\\
R^{SM}_g=3.435-0.172(g-r)_{\rm int},\nonumber\\
R^{SM}_r=2.638-0.077(g-r)_{\rm int},\nonumber \\
R^{SM}_i=1.819-0.02(g-r)_{\rm int},\nonumber\\
R^{SM}_z=1.383-0.017(g-r)_{\rm int},\nonumber\\
R_{tess}=1.902-0.179(g-r)_{\rm int},\label{eqn:vectors}
\end{eqnarray}

where $(g-r)_{\rm int}$ is the intrinsic color. With these color vectors we can both determine and account for extinction within a field anywhere on the sky. 

Since the extinction for each field is different, we must calculate it for each cutout. We calculate the extinction using stellar locus regression (SLR) with PS1 and SM photometry. For stars that lie on the stellar locus, extinctions act by shifting the stellar locus colors according to the extinction vector coefficients. In \tessreduce{}, we follow \citet{Tonry2012} and construct a PS1 and SM canonical stellar loci from synthetic photometry of the Calspec sources. In $r-i$ and $g-r$ space we then de-redden the observed PS1/SM magnitudes according to \autoref{eqn:extinction}. We fit the extinction by minimizing the distance of all sources to the canonical stellar locus. We limit the influence of non-stellar/peculiar sources by performing a 3$\sigma$ cut on the data and refit the extinction. 

\begin{figure}
    \centering
    \includegraphics[width=\columnwidth]{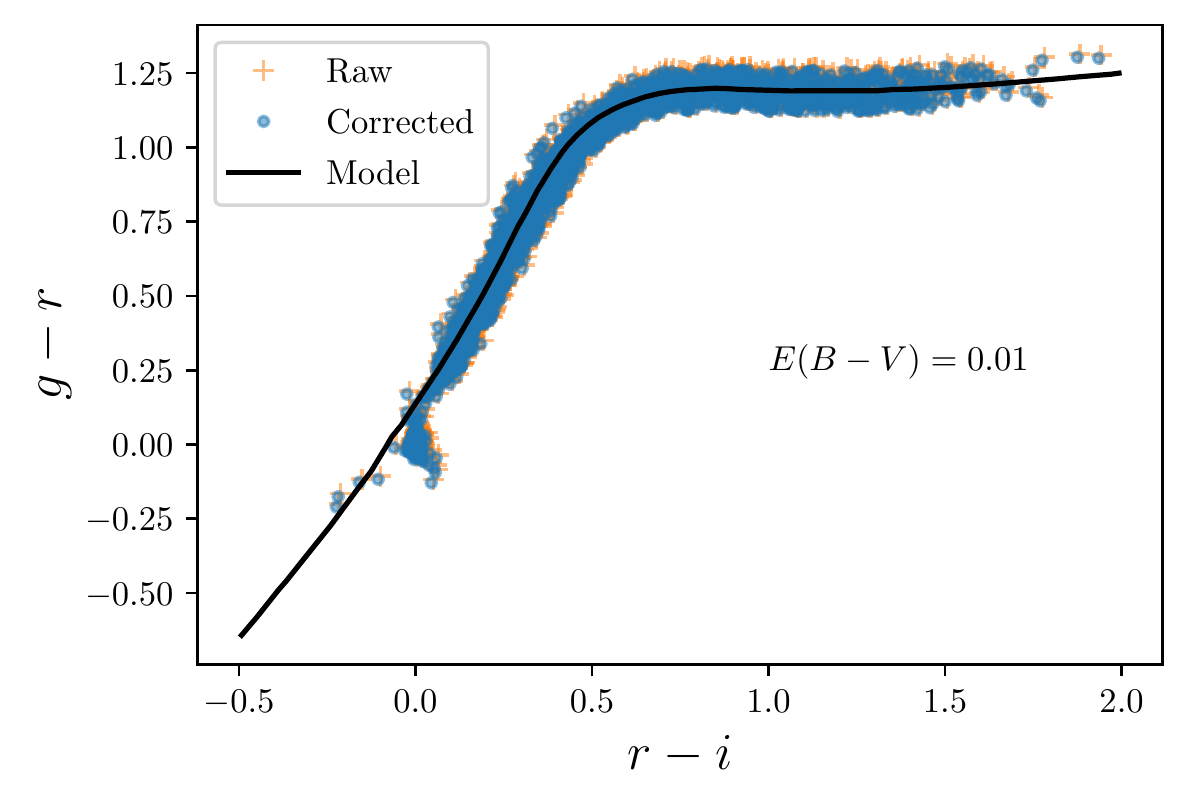}
    \includegraphics[width=\columnwidth]{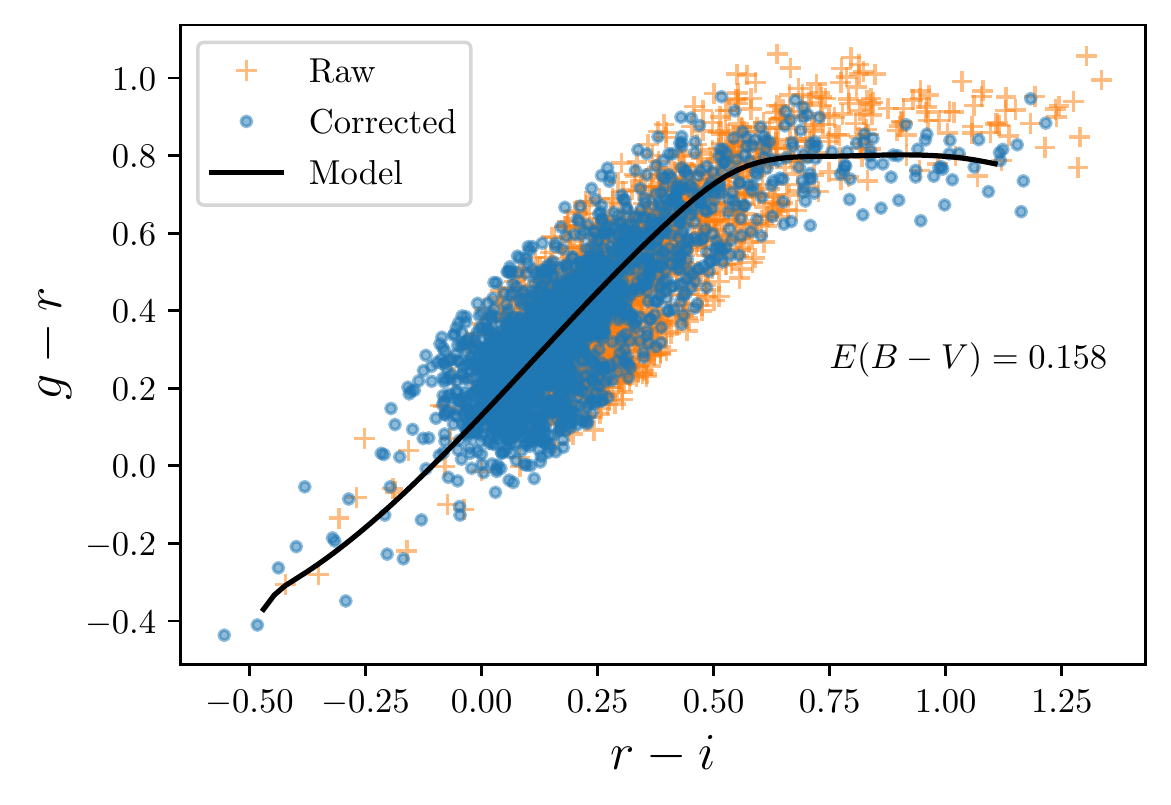}
    \caption{Stellar Locus Regression (SLR) for stars in the region surrounding two SN~2020cdj (\textbf{Top}) and SN~2020yzo (\textbf{Bottom}). To calculate $E(B-V)$ we shift the observed stellar locus according to the coefficients shown in \autoref{eqn:vectors} to de-redden sources until they match the model stellar locus. Since SN~2020cdj is in the North we use PS1 photometry, and SM photometry for the Southern SN~2020yzo.}
    \label{fig:slr_example}
\end{figure}

Since we use all PS1/SM DR1 catalog sources within the \textit{TESS} image, the stellar locus is well sampled in all fields. Example fits for dust extinction are shown in \autoref{fig:slr_example}. The extinction derived from SLR can then be incorporated into the construction of the composite \textit{TESS} magnitudes using the above extinction vectors to minimize the impact dust extinction has on instrument calibration. 

\subsection{In-situ zeropoint determination}

With the conversion from PS1/SM to \textit{TESS} magnitudes defined and a method for accounting for dust extinction, we are able to calculate a precise \textit{TESS} zeropoint across the entire sky. To calculate the zeropoint, we only use stellar sources that satisfy the following strict criteria:

\begin{itemize}
    \item $-0.5<(g-r)<1$,
    \item $14<i<17$,
    \item Isolated.
\end{itemize}

We make a cut on the source color to exclude red sources for which we have few Calspec sources. The color and magnitude conditions are determined using the PS1/SkyMapper DR1 photometry, however, we use the \textit{TESS} images to determine if the source is isolated. For each source we apply an annulus aperture, with an inner and outer radius of 6 and 7 pixels, respectively, to the reference image and sum all pixels. We calculate the magnitude of annulus aperture using a default zeropoint of 20.44 and require the annulus magnitude to be at lease 1 magnitude fainter than the \textit{TESS} magnitude we predict from PS1/SM photometry. If the annulus magnitude does not meet this condition we count the source as crowded and neglect it from the \textit{TESS} photometric calibration. This provides an easy check to determine if the source is crowded by other nearby bright sources, that might be ignored by a simple distance measure. An alternative method would be to use source catalogs to identify separation, however, this offers its own challenges with adopting a variable separation distance check for sources based on their magnitudes and will struggle to account for the contamination of large extended sources, like galaxies. Our calibration method assumes that the majority of sources within the magnitude range of $14<i<17$ are well contained in a $3\times3$~pixel aperture, which holds for stellar sources. Sources that pass this selection are then used in determining the zeropoint.

With a selection of high quality sources we can reliably determine the \textit{TESS} photometric zeropoint. First, we calculate the \textit{TESS} instrumental magnitude for all sources via $m_{inst}=-2.5log(C)$, where $C$ is the measured \textit{TESS} counts. We then assert that this instrumental magnitude equals the composite \textit{TESS} magnitudes constructed from PS1 photometry plus a zeropoint. From this assertion the zeropoint is found simply as:

\begin{eqnarray}
zp = m_{com} + 2.5log(C).
\end{eqnarray}

\begin{figure*}
    \centering
    \includegraphics[width=\textwidth]{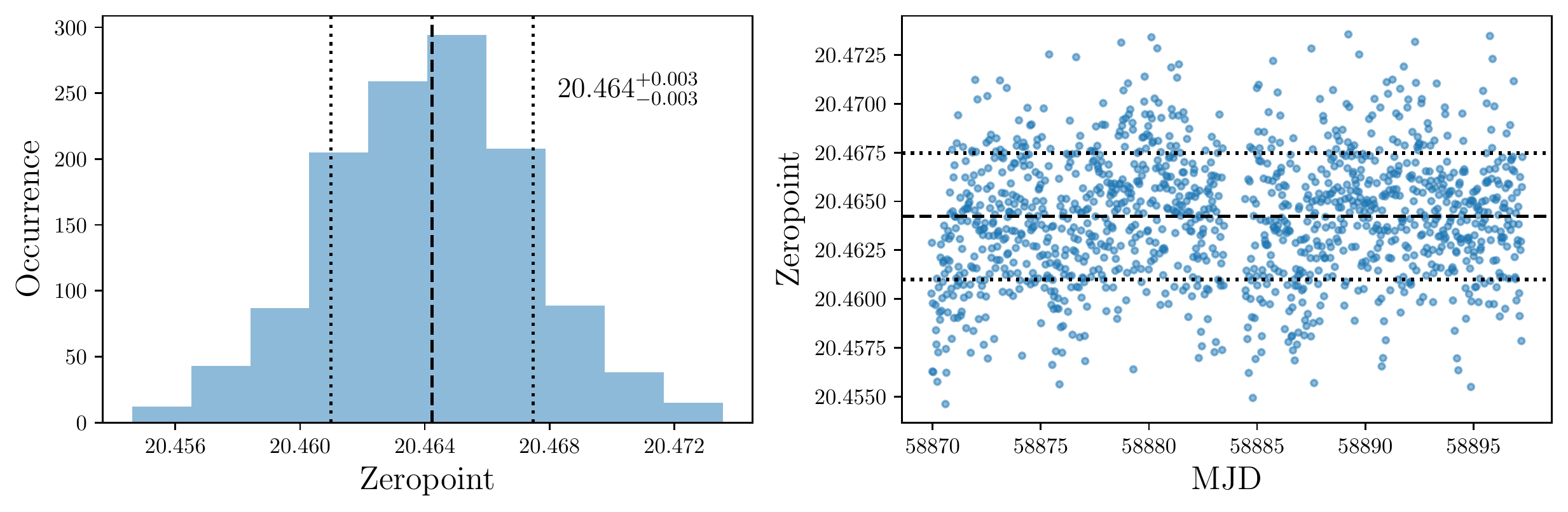}
    \caption{Zeropoint diagnostics for the region containing SN 2020cdj, which achieves a very well defined value. In both figures the black dashed line is the median, while the black dotted lines are the $16^{th}$ and $80^{th}$ percentiles. These values are also shown on the left panel. \textbf{Left:} Cumulative distribution of all zeropoint measurements across a sector. \textbf{Right:} Zeropoints through time for all images in a sector. Generally we find the \tessreduce{} zeropoints are well constrained and are larger than the theoretical zeropoint of 20.44. Therefore, using the theoretical zeropoint will overestimate the brightness of objects according to \textit{TESS}. We calculate this zeropoint by comparing $3\times 3$~pixel aperture photometry of isolated stars with \textit{TESS} composite magnitudes derived from PS1 observations. This data driven zeropoint can be precise and offers a post processing correction to systematics in the light curves by letting the zeropoint vary with time.}
    \label{fig:zp_example}
\end{figure*}

With this simple equation we can calculate a data driven zeropoint for each image in the sector. This time series of zeropoints can then be used directly to provide a post reduction correction to the light curve, or all zeropoints can be averaged to achieve a high precision zeropoint. In many cases, such as seen in \autoref{fig:zp_example}, the zeropoint converges to percent precision.

Generally, we find that the average \textit{TESS} zeropoint of 20.44 from \citet{TESShandbook} is lower than the zeropoints derived by \tessreduce{}. While we have yet to fully explore the nature of the \textit{TESS} zeropoint, we are confident that this method produces accurate zeropoints. \citet{Ridden-Harper2021b} shows that this calibration method is consistent with two independent flux calibration methods for the \textit{Kepler} broadband filter. 

This powerful method allows us to produce reliable flux calibrations for \textit{TESS} light curves wherever PS1 or SM photometry is available. In future work, we will explore how the \textit{TESS} zeropoint varies between detectors across the entire focal plane.

\section{Reduction pipeline} \label{sec:reduction}

We developed \tessreduce{} to be a user-friendly and reliable reduction pipeline to allow anyone to use \textit{TESS} data. We have streamlined \tessreduce{} such that a user with no knowledge of the intricacies of \textit{TESS} systematics could extract light curves. Here we will present the main reduction methods and elaborate on the pipeline structure. 

Due to the sector observing plan, it is often not clear if \textit{TESS} observed a given transient. To make it easy to check if observations exist for a certain target, we have included functions to check if a target was covered by \textit{TESS}. For transients with catalog names, the \texttt{sn\_lookup} function can be used. This function utilizes the OSC API to retrieve objects' coordinates and key times, such as discovery (disc) and maximum (max) epochs. We then check to see what sectors the retrieved coordinates were observed by \textit{TESS} with \texttt{tess-point} \citep{tesspoint2020}, and then check to see if the selected time is covered by a \textit{TESS} sector, and if not what the time difference is. An example output table output by the above command is shown in \autoref{tab:sn_lookup}. 

Alternatively, if the object is not in a transient catalog, then the \texttt{spacetime\_lookup} function can be used. For this instance, we compare input coordinates and time in MJD to \texttt{tess-point} and the sector time list. The output of both of our observation checking functions can be directly input into \tessreduce{} to download and reduce overlapping data. We give an example of the reduction process using the search functions as inputs in \autoref{sec:code_example}.

Generally, \tessreduce{} can take the following inputs to act on:
\begin{itemize}
    \item direct input of a TPF,
    \item RA, Dec. and sector,
    \item output from the \tessreduce{} observation lookup functions.
\end{itemize}
Unless a TPF is directly provided \tessreduce{} will query \textit{TESS}cut through Lightkurve functions to obtain the relevant TPF. \textit{TESS}cut restricts the total size for any downloaded TPF to be $<100$~MB.

In each case, \tessreduce{} will extract the relevant information needed for data reduction. Unless the user sets \texttt{reduce=False} in the class, then the data will automatically be reduced via the difference image pathway, flux calibrated and aperture photometry performed with a $3\times 3$~pixel square aperture. 

Individual steps of the reduction can be disabled. Such steps include image alignment, photometric calibration, and difference imaging. Any combination of these options can be enabled or disabled, however, the default option of all enabled produces the best result. In the complete reduction the process is as follows:
\begin{itemize}
    \item Obtain reference image
    \item Generate an image mask
    \item Calculate and subtract background
    \item Calculate spatial shifts for each image, relative to the reference$^\dagger$
    \item Reset flux to the raw \textit{TESS} images
    \item Align images according to the calculated shifts
    \item Subtract reference image
    \item Recalculate and subtract background
    \item Calibrate photometry with field stars$^\dagger$
    \item Calculate target light curve with aperture photometry$^\dagger$
\end{itemize}

\begin{figure*}
    \centering
    \includegraphics[width=\textwidth]{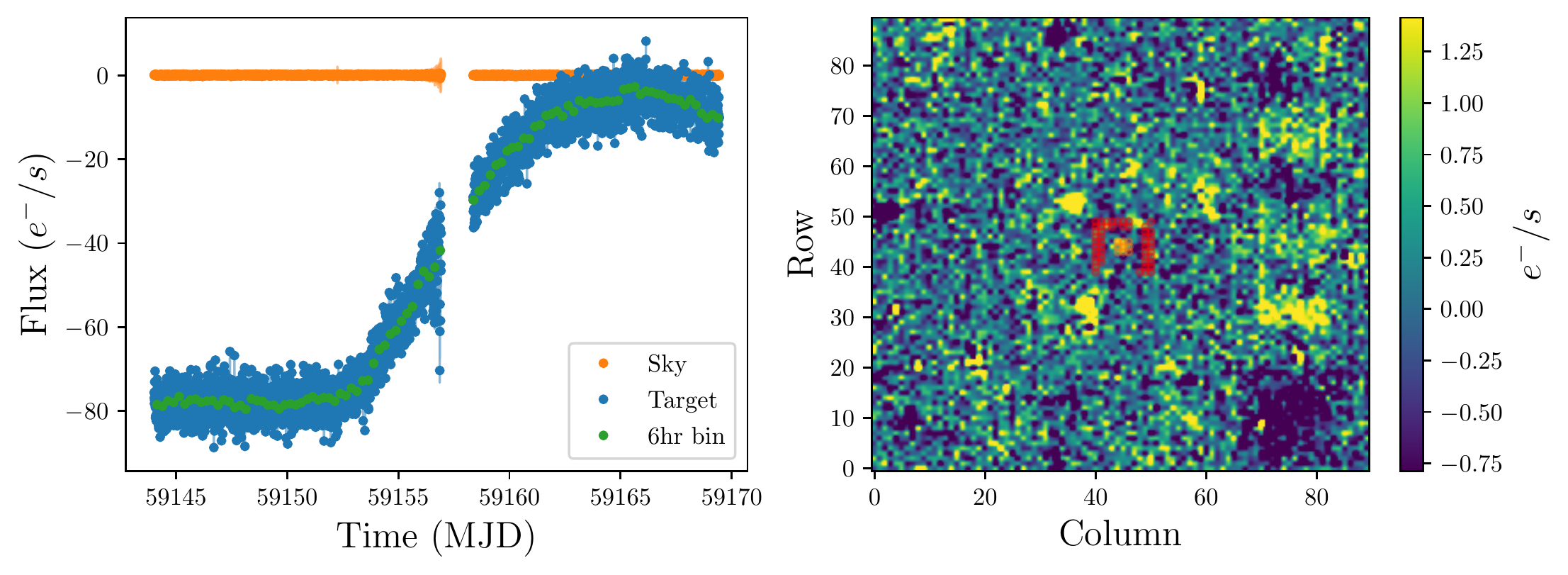}
    \caption{Reduction quick look figure produced by the \tessreduce{} pipeline for SN~2020yzo, a spectroscopically classified SN~II. \textbf{Left:} light curve for SN~2020yzo calculated with aperture photometry and the average sky background. \textbf{Right:} Difference image of the $90\times90$~pixel field containing the SN.}
    \label{fig:diff_diag}
\end{figure*}

Key steps in the reduction, marked by $\dagger$, will produce diagnostic plots if plotting is enabled. These diagnostic plots include \autoref{fig:align}, \autoref{fig:slr_example}, \autoref{fig:zp_example} and the final reduction in \autoref{fig:diff_diag}. The final reduction figure shows the target light curve alongside the local sky light curve on the left and an example difference image on the right for the frame that contains the highest flux. These plots are a great quick check for the reduction quality.

Following the reduction of \textit{TESS} data with \tessreduce{}, there are additional functions to aid in the analysis. For example, there are functions to easily convert from \textit{TESS} counts to magnitudes or flux. When converting to flux, users can chose between the following flux units: Jy, mJy, cgs. The \textit{TESS} zeropoint is tracked to allow for easy shifts.

For transients that span an entire \textit{TESS} sector, the reference image will always include the transient flux, leading to an overall offset in the light curve, as seen in \autoref{fig:reduction_example}. Therefore, deriving an accurate host subtraction can be challenging and is one of the limitations of \tessreduce{}. However, future work will focus on predicting a \textit{TESS} scene from ground based data to construct accurate, host-subtracted images.

Alongside the \textit{TESS} light curve, \tessreduce{} also has a simple aggregate for light curves from other telescopes. Within \texttt{ground} we collect supplementary light curves from public databases. We access the public ZTF $g$ and $r$ light curves through the ALeRCE API. This data can easily be compared to the \textit{TESS} light curve by using the \texttt{plotter} function. This comparison can provide key insights into the nature of objects observed by \textit{TESS} and as a way to verify the \textit{TESS} light curve.

For studies of stellar sources, Lightkurve offers key tools in light curve analysis. To make use of all Lightkurve functions, the \tessreduce{} light curve can be converted to the relevant Lightkurve format with a single function that carries the flux/counts units.

Examples of the analysis steps outlined here can be found on Github for a SN\footnote{\url{https://github.com/CheerfulUser/TESSreduce/blob/master/SN_Example.ipynb}} and a variable star\footnote{\url{https://github.com/CheerfulUser/TESSreduce/blob/master/Var_example.ipynb}}.

\section{Example light curves} \label{sec:light curves}

\begin{figure}
    \centering
    \includegraphics[width=\columnwidth]{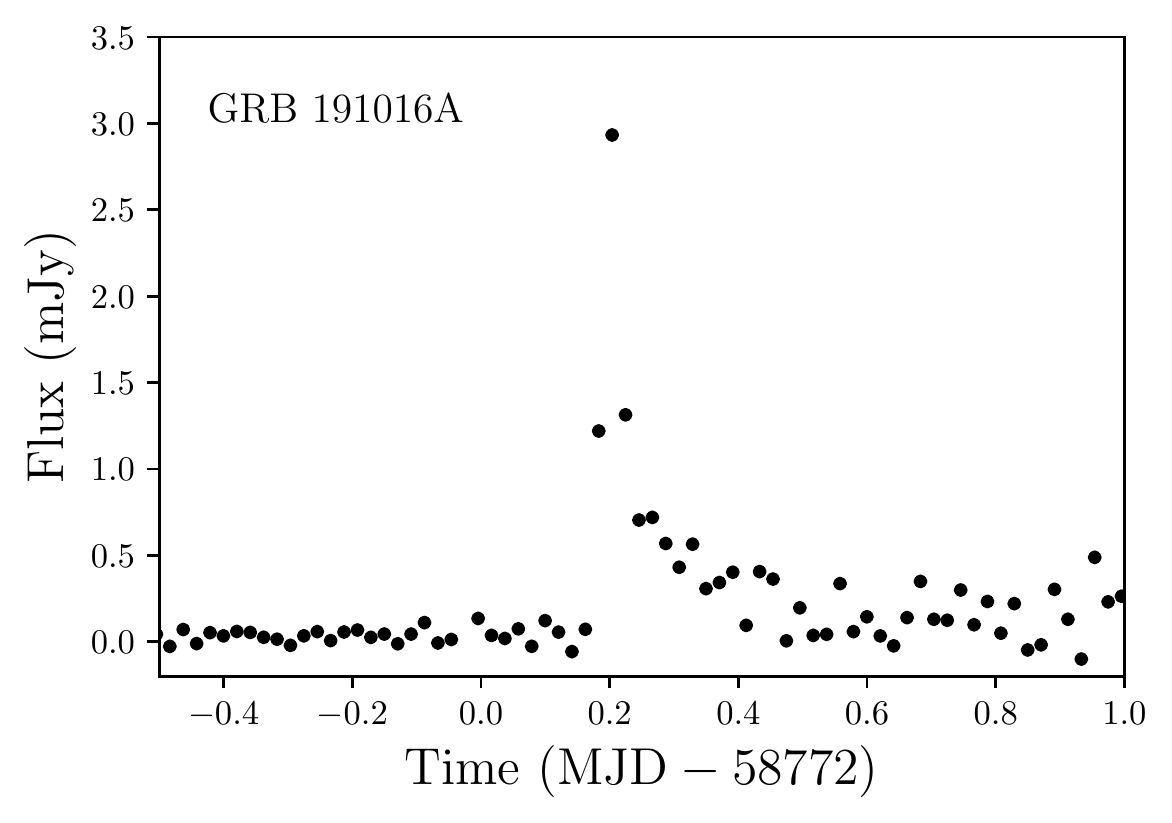}
    \includegraphics[width=\columnwidth]{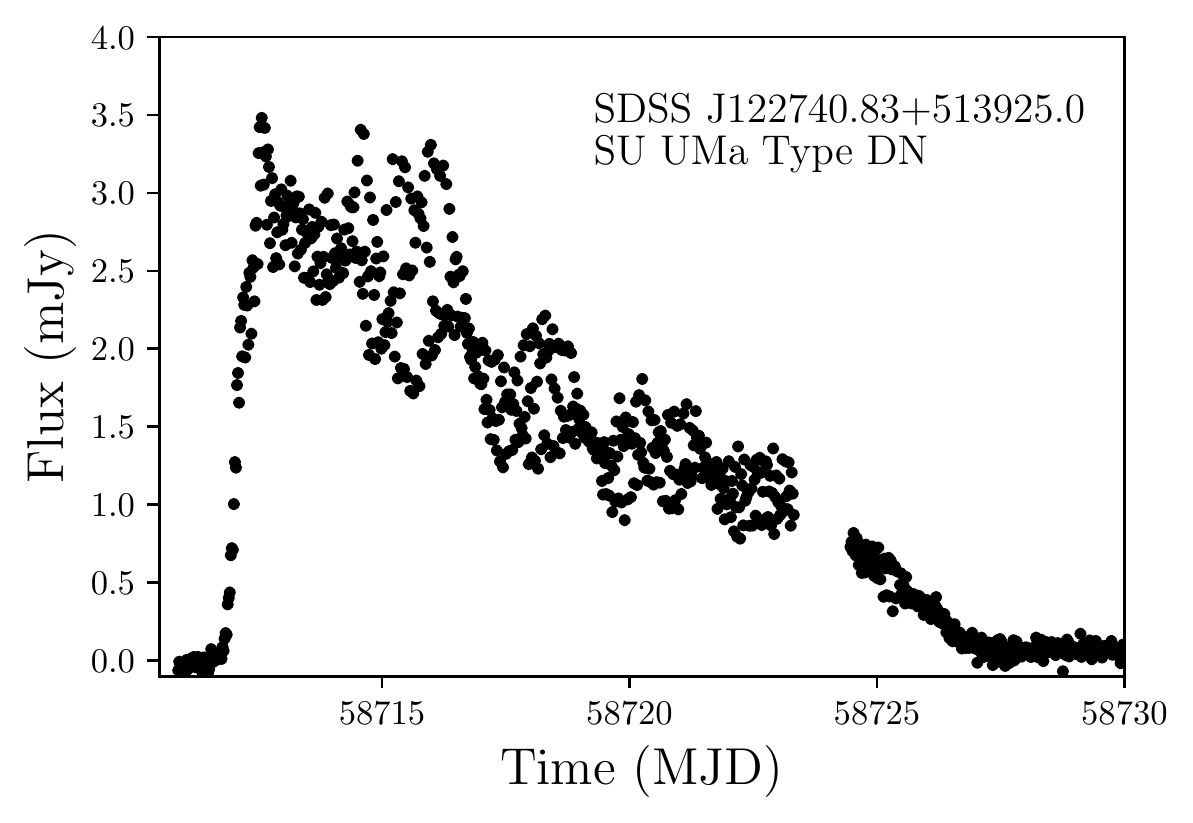}
    \includegraphics[width=\columnwidth]{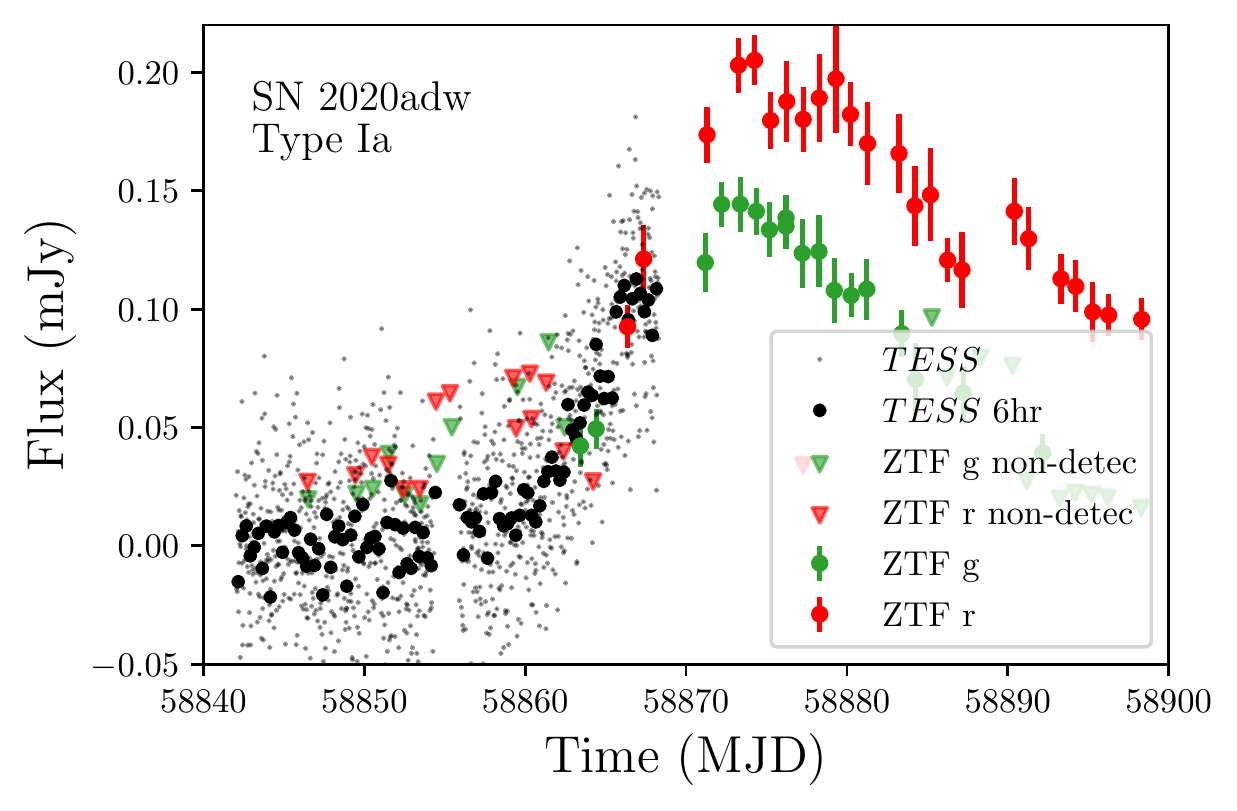}
    \caption{A selection of light curves reduced with and calibrated \tessreduce{}. \textbf{Top}: A bright optical afterglow of GRB 191016A, while rare, \textit{TESS} is expected to observe 1-2 events each year \citep{Smith2021}. \textbf{Middle}: A super-outburst of SDSS J122740.83+513925.0, a SU UMa type dwarf nova in 2019, from this data system parameters such as superhump period can be well constrained. \textbf{Bottom}: The rise of the type Ia supernova SN~2020adw, alongside the public ZTF data \citep{sn2020adw}. These examples are a tiny fraction of objects that have been observed by \textit{TESS} and can be readily reduced with \tessreduce{}.}
    \label{fig:lc_examples}
\end{figure}

Since \tessreduce{} is a general forced photometry pipeline, the methods are applicable to all objects. To demonstrate the versatility of this pipeline, we show a selection of light curves from a variety of objects in \autoref{fig:lc_examples}. In the top panel, we show the light curve of GRB~191016a, a bright GRB afterglow observed by \textit{TESS}, which was presented in \citet{Smith2021}. While such events are relatively rare, \citet{Smith2021} predicts that \textit{TESS} should observe $1-2$ such events each year. \tessreduce{} is an ideal tool to rapidly check analyze short and unique transients in \textit{TESS} data. 

Longer duration events, such as dwarf novae, and novae outbursts are also readily accessible with \tessreduce{}. \autoref{fig:lc_examples} (\textbf{middle}) shows a super-outburst from the SU UMa type dwarf nova SDSS J122740.83+513925.0 in 2019 that was observed by \textit{TESS}. From this light curve both the super-hump period and orbital period can be identified as $P_1=0.06464$~d and $P_{orb}=0.06295$~d respectively. These values are consistent with those presented in \citet{Kato2012} from the 2007 ($P_1= 0.064604$~d, $P_{orb}=0.062950$~d) and 2011 ($P_1= 0.064883$~d, $P_{orb}=0.062950$~d) superoutbursts, and may indicate evolution of the system.

As has been seen already, \tessreduce{} is capable of supernovae reduction. \autoref{fig:lc_examples} (\textbf{bottom}) shows the rise of SN~2020adw, a type Ia supernova, alongside the public ZTF data. The close comparison to ZTF photometry showcases the flux calibration method used in \tessreduce{}, which is also seen in \autoref{fig:reduction_example} for SN~2020cdj. From this example, it is also clear that under ideal circumstances, e.g., well behaved background and minimal crowding, \textit{TESS} can provide unparalleled high precision, high cadence light curves for supernovae at early times, with a detection limit surpassing that of ZTF. Such data could provide valuable insights into the nature of supernovae. 

The examples shown here are just a limited suit of test cases, numerous targets can already be reduced with the current \tessreduce{} functions, however, we welcome community engagement to help build additional functionality to meet other science goals.

\section{Conclusions}

\textit{TESS} presents an incredibly valuable data set to study the time evolution of transients and many other time varying phenomena. The high cadence imaging can capture key moments that could provide essential clues into the progenitors of supernovae, the nature of fast transients, such as GRBs, RETs, and flare stars, and offers the exciting possibility of searching for new transients in the short time domain.  Challenges with data reduction generates an entry barrier to the use of \textit{TESS} data. These challenges stem primarily from the highly variable scattered Earth/Moon light background, apparent telescope motion and flux calibration. Here we have presented the \tessreduce{} package which allows for easy \textit{TESS} data reduction, accounting for the key challenges with \textit{TESS} data. 

Unlike the other reduction pipelines \tessreduce{} leverages data from other observatories. Using the PS1, SkyMapper, and Gaia photometric catalogs, with accurate and complete photometric catalogs to the depth of \textit{TESS}, allow us to a priori construct source masks and estimate the \textit{TESS} magnitudes of targets. This approach minimizes biases that arise from including sources in the background, that other difference imaging methods will encounter. Furthermore, our approach to accounting for effect of straps is physically motivated, unlike methods used in existing pipelines. 

While \textit{TESS} can provide precise relative photometry, it must be calibrated to physical flux to compare against theoretical models and other observations. Variations of CCD detectors means that a single zeropoint is insufficient to characterize all \textit{TESS} data. With ground based $grizy$ photometry we can reliably predict \textit{TESS} magnitudes to determine the zeropoint locally based on field stars. To have robust estimated magnitudes, we account for dust extinction, and constrain how extinction impacts the \textit{TESS} filter. With this novel calibration method, \tessreduce{} can provide precise zeropoints that agree well with independently calibrated data.

To reduce complexities in \textit{TESS} data analysis, we include a number of functions in \tessreduce{} that make identifying and using \textit{TESS} easier. These functions include observation lookup functions and a way to directly compare to ground based public ZTF data. These functions make checking for data and comparing against independent data as simple as a single Python command. 

We demonstrate the versatility of \tessreduce{} by examining a GRB optical afterglow, dwarf nova superoutburst, and supernovae. These examples only show a fraction of the variety of variable objects that have been observed by \textit{TESS}, and can be reduced with \tessreduce{}. We have developed \tessreduce{} to make reducing \textit{TESS} data simple, and easy for anyone to use without needing pre-existing knowledge of \textit{TESS} data collection and systematic. As \tessreduce{} is an open source package, we encourage users to report issues and lodge pull requests to help improve the capabilities of \tessreduce{} and the quality of data reduction.

\section*{Acknowledgments}

\textit{Software}: \\
\texttt{numpy} \citep{numpy}, 
\texttt{matplotlib} \citep{matplotlib}, 
\texttt{pandas} \citep{pandas},
\texttt{scipy} \citep{scipy},

This research made use of Lightkurve, a Python package for \textit{Kepler} and \textit{TESS} data analysis (Lightkurve Collaboration, 2018).
This work has made use of data from the European Space Agency (ESA) mission
{\it Gaia} (\url{https://www.cosmos.esa.int/gaia}), processed by the {\it Gaia}
Data Processing and Analysis Consortium (DPAC,
\url{https://www.cosmos.esa.int/web/gaia/dpac/consortium}). Funding for the DPAC
has been provided by national institutions, in particular the institutions
participating in the {\it Gaia} Multilateral Agreement.
This research has made use of the SVO Filter Profile Service (http://svo2.cab.inta-csic.es/theory/fps/) supported from the Spanish MINECO through grant AYA2017-84089. The material is based upon work supported by NASA under award number 80GSFC21M0002.
TMB was funded by the CONICYT PFCHA / DOCTORADOBECAS CHILE/2017-72180113.

\section*{Data Availability}
All data we make use of is publicly available via MAST, Vizier, ALeRCE, and the Open Supernova Catalog. This data can be easily retrieved through routines built into \tessreduce{}.



\bibliographystyle{mnras}
\bibliography{AllPhD} 




\appendix

\section{Image alignment vs. Delta kernels} \label{sec:shift_vs_delta}
One key difference between \tessreduce{} and other difference imaging pipelines is that \tessreduce{} does not perform kernel matching. While necessary for ground based telescopes which suffer from large PSF variations between observations, \textit{TESS} has a relatively stable PSF, and therefore PRF. Other pipelines such as ISIS based pipelines and the DIA pipeline use kernel matching with Delta kernels, otherwise known as discrete kernels, to correct for image alignment and any small changes in the PRF. As described in \citet{Miller2008} a Delta kernel, $K$, is constructed from an orthogonal basis of vectors $K_n$ as follows:
\begin{eqnarray}
K=\sum_{n=1}^N a_n K_n,
\end{eqnarray}
where $a_n$ is the vector coefficient. In principle any given \textit{TESS} image should be related to another image through a convolution with an appropriate Delta kernel. 

Normally the coefficients to the basis which constructs the Delta kernel are derived by simultaneously fitting to a collection of isolated stars across the detector, however, with the relatively small cutout used by \tessreduce{} we instead fit the coefficients by minimising the difference between two images. Following a similar process to how we derive the image shifts, we first calculate and subtract the background from every image. With these background subtracted images we then use the \texttt{scipy} \texttt{minimize} algorithm to minimize the absolute difference in counts between a given image and a reference image convolved with a $7\times7$~pixel kernel using the \texttt{scipy} \texttt{fftconvolve} algorithm. To avoid edge effects we ignore the boundary up to 7 pixels in the minimization. This process yields a time series of Delta kernels that when convolved with the reference image give the best fit to the data. To generate a difference image, we subtract the convolved reference image from each frame, with which we then calculate and subtract the background.

\begin{figure*}
    \centering
    \includegraphics[width=\textwidth]{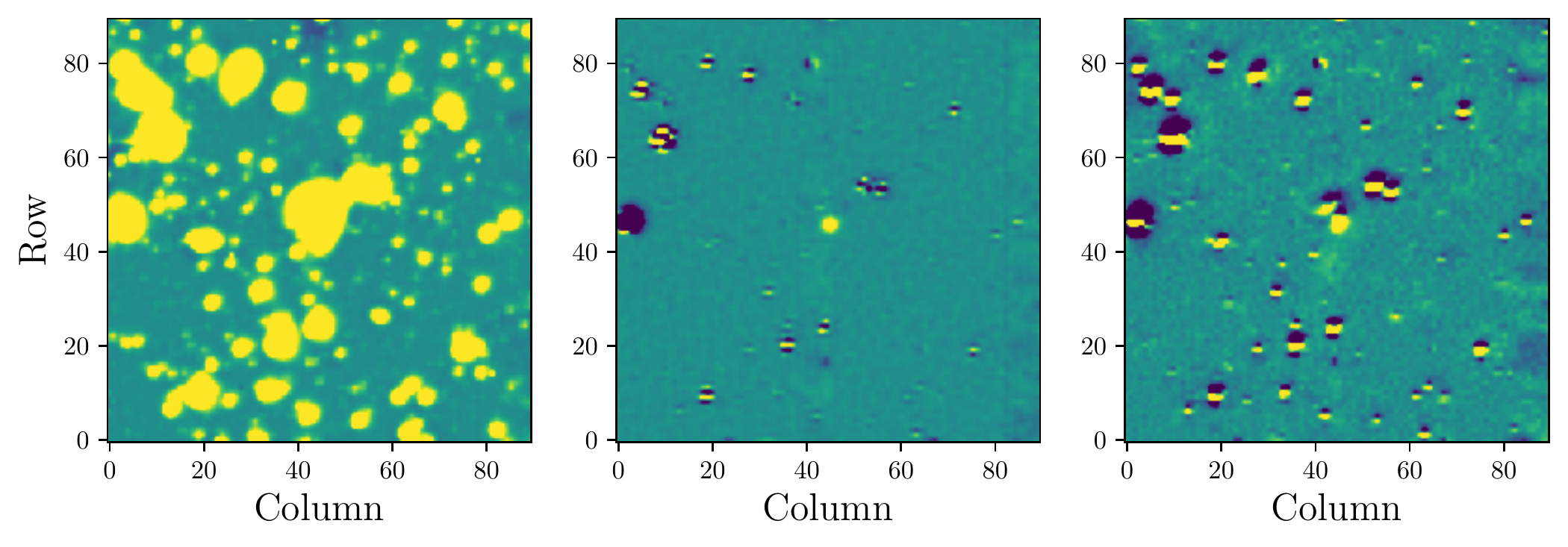}
    \caption{Comparison of subtraction methods for a $90 \times 90$~pixel FFI cutout centered on SN~2020fqv. This is a good test case since the host galaxy is resolved in \textit{TESS}, so correct alignment is crucial. \textbf{Left:} Background subtracted image. \textbf{Middle:} Difference image created by the \tessreduce{} image alignment, all sources are well subtracted and the SN is clearly visible. \textbf{Right:} Difference image created by kernel matching, while many sources are subtracted the kernel is not well defined from this data set, leaving significant residuals around the SN. Clearly for cutouts such as those used by \tessreduce{}, image shifting is the better approach. All images are on the same colour scale which spans $-10$ to $+10$ counts.}
    \label{fig:kernel}
\end{figure*}

With this process we now have a kernel matched difference image that we can compare to the image alignment method. As seen in \autoref{fig:kernel} we test the two methods on a \textit{TESS} FFI cutout centered on SN~2020fqv. In this field there are numerous stellar sources, and a large central galaxy, which contains SN~2020fqv (\textbf{Left}). We find that the image alignment method produces more reliable subtractions (\textbf{Middle}) than the kernel matching method (\textbf{Right}). Furthermore, the kernel matching method uses significantly more compute time ($\gtrsim 5\times$) than the image alignment method. From this analysis we conclude that image alignment is the best method for \tessreduce{}. 

\section{Comparing \textit{TESS} to Gaia G}
\begin{figure}
    \centering
    \includegraphics[width=\columnwidth]{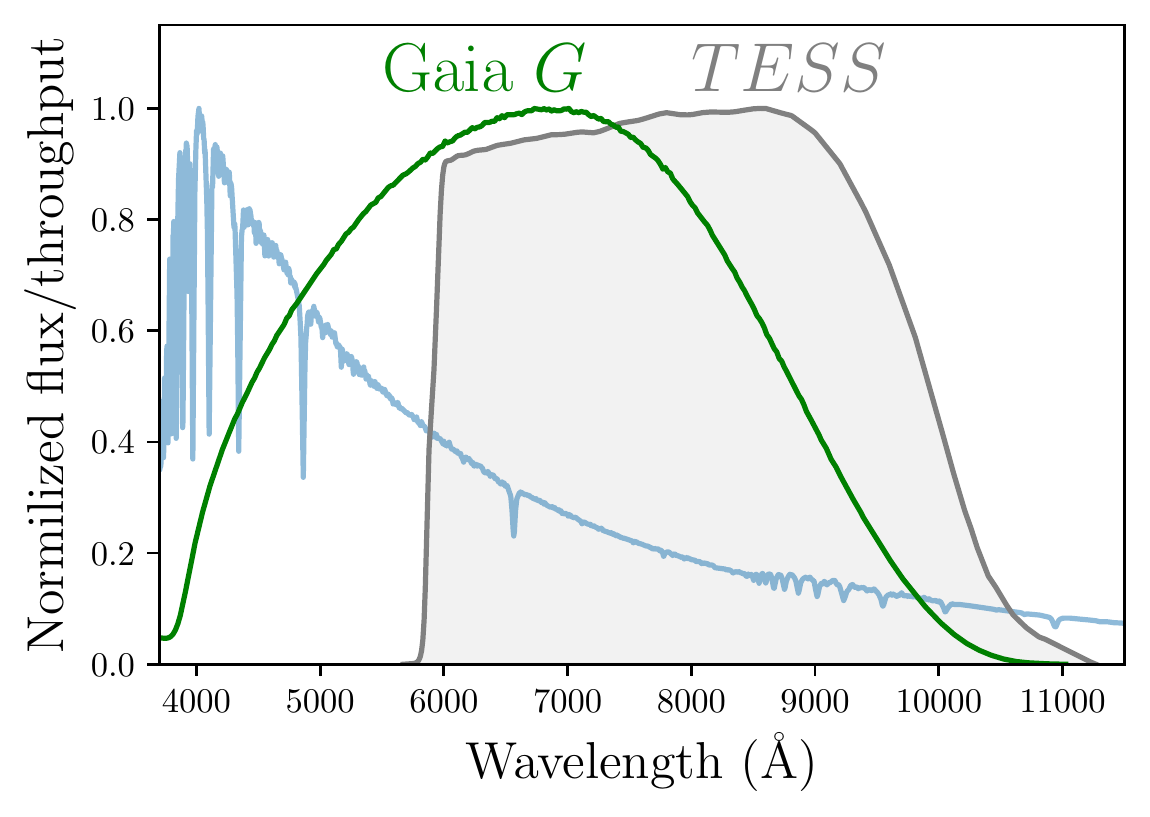}
    \caption{The Gaia $G$ and \textit{TESS} filters overlaid on the Calspec spectra for HD~2811. While Gaia $G$ overlaps with most of the \textit{TESS} filter, it also extends to much shorter wavelengths, making it unreliable for calibrating \textit{TESS} photometry against.}
    \label{fig:gaia_tess}
\end{figure}

While Gaia G does cover similar wavelengths to \textit{TESS}, it extends to notably shorter wavelengths. The difference can be sen in \autoref{fig:gaia_tess}. This large difference will introduce a strong color dependency in the mapping form Gaia to \textit{TESS} magnitudes. By using PS1 and SM color photometry to reconstruct the \textit{TESS} filter, we can achieve a much closer approximation of the true \textit{TESS} magnitude and account for any residual color terms.

\section{SkyMapper dust extinction vector coefficients}
\begin{figure*}
    \centering
    \includegraphics[width=\textwidth]{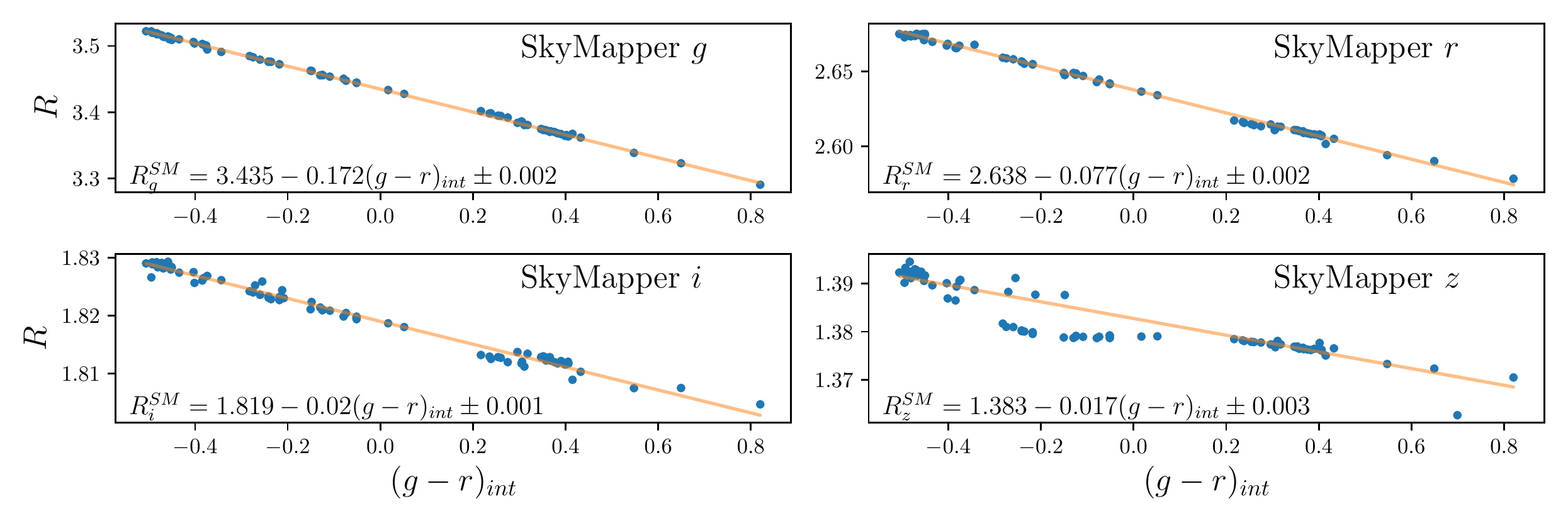}
    \caption{Extinction vector coefficients, $R$, for the SkyMapper $griz$ filters. The construction of these coefficients follows the same process used for the PS1 and \textit{TESS} filters. The equations for each fit alongside the 1$\sigma$ error.}
    \label{fig:R_skymapper}
\end{figure*}

Following the process described in \autoref{sec:dust} we construct the color dependent extinction vector coefficients for the SkyMapper $griz$ filters. The fits to the data can be seen in \autoref{fig:R_skymapper}.

\section{Example reduction} \label{sec:code_example}
In this section we have an example reduction for SN~2020cdj, as seen in \autoref{fig:reduction_example}. To check that a target has been observed by \textit{TESS}, the \texttt{sn\_lookup} function can be used to provide a quick check. This function produces a table as shown in \autoref{tab:sn_lookup}, and the output of this function can be used as an input for \tessreduce{}.

\begin{table}
 \caption{Example of the table output by \texttt{sn\_lookup()} for discovery time of SN~2020cdj. Sectors before 17 and after 25 are excluded here for brevity.}
 \centering
 \label{tab:sn_lookup}
 \begin{tabular}{lcc}
  \hline
  Sector & Covers & Time difference \\
         &        & (days)\\
  \hline
    17 & False & -98 \\
    18 & False & -73 \\
    20 & False & -18 \\
    21 & True & 0 \\
    22 & False & 10 \\
    23 & False & 39 \\
    24 & False & 68 \\
    25 & False & 95 \\
  \hline
 \end{tabular}
\end{table}

\begin{verbatim}
    import tessreduce as tr
    obs = tr.sn_lookup('sn2020cdj')
    tess = tr.tessreduce(obs_list=obs)
    tess.to_flux()
    tess.plotter(ground=True)
\end{verbatim}
With these few lines of code \tessreduce{} will download the \textit{TESS} data, reduce and calibrate the data, then plot it alongside the public ZTF light curves. The final plotter command will produce a figure similar to the right panel of \autoref{fig:reduction_example}.

If the object is not a transient listed on the OSC, then overlapping observation can be found by using the \texttt{spacetime\_lookup()} function. This function takes arguments of R.A. Dec. and time in MJD. If a TPF is already downloaded, the TPF can be input directly as \texttt{tr.tessreduce(tpf=FILE)}.

Following the reduction, difference imaged lightcurves can be produced at any position in the image with \texttt{tess.diff\_lc()}, where the inputs are the $x$, and $y$ pixel positions of the aperture center. To analyse the lightcurve using the extensive tools made available with the Lightkurve package, then the output lightcurve can be converted the a LightCurve object with \texttt{tess.to\_lightkurve()}.

Further information can be found in the docs\footnote{\url{https://tessreduce.readthedocs.io/en/latest/}}, and example reductions for a supernova\footnote{\url{https://github.com/CheerfulUser/TESSreduce/blob/master/SN_Example.ipynb}} and a variable star\footnote{\url{https://github.com/CheerfulUser/TESSreduce/blob/master/Var_example.ipynb}} can be found on the Github.


\bsp	
\label{lastpage}
\end{document}